\documentclass[aps,prd,reprint,twocolumn,amsmath,amssymb,superscriptaddress,longbibliography,floatfix]{revtex4-2}
\usepackage[T1]{fontenc}
\usepackage{graphicx}
\usepackage{bm}
\usepackage{booktabs}
\usepackage{placeins}
\usepackage{needspace}
\usepackage{afterpage}
\usepackage{subfiles}
\usepackage{float}
\usepackage{capt-of}
\usepackage{hyperref}
\usepackage{orcidlink}
\newcommand{\RIDDLE}{\textsc{RIDDLE}}
\newcommand{\mjj}{m_{jj}}
\newcommand{\epsB}{\epsilon_{\mathrm B}}
\newcommand{\epsS}{\epsilon_{\mathrm S}}
\newcommand{\SIC}{\mathrm{SIC}}
\newcommand{\SR}{\mathrm{SR}}

\newcommand{\zb}{\bm{z}}
\newcommand{\xb}{\bm{x}}
\newcommand{\dd}{\mathrm{d}}

\newcommand{\TeV}{\,\mathrm{TeV}}
\newcommand{\GeV}{\,\mathrm{GeV}}

\hypersetup{hidelinks,pdftitle={Revealing irregularities through distributional discrepancies in latent space via Stein-witness estimation},pdfauthor={Ali Eren Simsek}}
\newcommand{\plotpanel}[2]{%
  \begin{minipage}[t]{0.485\linewidth}%
    \centering
    \includegraphics[width=\linewidth]{#1}%
    \if\relax\detokenize{#2}\relax
    \else
      \par\smallskip
      {\small #2\par}%
    \fi
  \end{minipage}%
}

\newcommand{\resultsectionbreak}{%
    \newpage
    \null\newpage
    \null\newpage
}

\begin{document}
\title{Revealing irregularities through distributional discrepancies in latent space via Stein-witness estimation}
\author{Ali Eren Simsek\,\orcidlink{0000-0002-9074-2256}}
\email{ali.eren.simsek@cern.ch}
\affiliation{Department of Physics, The Catholic University of America, Washington, DC, USA}

\hyphenpenalty=10000
\exhyphenpenalty=10000
\emergencystretch=2em


\begin{abstract}
\RIDDLE{} is a new method for resonant anomaly detection that uses a scalar Stein-witness function to identify discrepancies between unlabeled signal-region events and a background distribution inferred from dijet-mass sidebands. The witness is optimized in a latent representation of auxiliary jet observables against a differentiable background reference, without requiring an explicit signal model. The method is evaluated using simulated LHC Olympics dijet events and compared with its idealized and supervised benchmarks, which share the same statistical construction but use additional background or signal information, and with LaCATHODE as an independent anomaly detection method. \RIDDLE{} demonstrates strong signal--background discrimination comparable to the idealized benchmark, generally outperforming LaCATHODE while largely preserving the background dijet-mass spectrum under anomaly-score selection. Tests incorporating additional angular information and deliberately mass-correlated features further demonstrate the method's robustness under the studied conditions. Together, these results highlight the potential of \RIDDLE{} as a robust, data-driven approach for enhanced resonance searches that combines effective anomaly discrimination with background-shape preservation.
\end{abstract}

\maketitle
\raggedbottom

\section{Introduction}\label{sec:introduction}

Searches for narrow resonances in the dijet invariant mass, $\mjj$, provide a broadly applicable probe of new particles. Their sensitivity is limited by the large, smoothly falling quantum chromodynamic (QCD) multijet background. Auxiliary observables, such as individual jet masses and jet substructure, can help distinguish signal-like events from ordinary multijet production. However, correlations between these observables and $\mjj$ can reshape the background mass spectrum through event selection, potentially creating structures that resemble a resonance. An enhanced bump hunt must therefore improve signal--background separation while preserving a background shape suitable for a subsequent resonance search. Background estimation from mass sidebands and discrimination in background-trained latent representations provide established approaches to this problem~\cite{collins2018cwola,nachman2020anode,hallin2022cathode,hallin2023lacathode}.

\RIDDLE{} is introduced as a new method for resonant anomaly detection based on a scalar \emph{Stein-witness function}, rather than a binary classifier or an explicit signal-density model. A background reference is inferred from dijet-mass sidebands, the regions outside a candidate signal region. Here, the witness is a neural network with a single scalar output, trained using unlabeled signal-region events in a learned latent representation of auxiliary observables. Its principal training objective uses the witness derivatives to expose distributional differences rather than to predict event labels, following the principle of learned Stein critics~\cite{grathwohl2020stein,repasky2023critic}. To account for changes in the background across the dijet-mass spectrum, the mass conditions the background mapping and density model, while the witness itself receives only the resulting latent auxiliary coordinates. The witness derivatives, together with the background reference, define a \emph{Stein response}. Stein's identity states that the mean response under the background reference is zero, provided suitable smoothness and boundary conditions hold~\cite{stein1972bound,grathwohl2020stein}. This is an expectation for the background distribution, not a requirement that every event have zero response; the average over a finite sample can also differ from zero because of statistical fluctuations. The witness is trained to reveal differences between the signal-region data and the background reference by looking for departures from this zero-mean expectation. Its output and derivatives are then used to construct an anomaly score for each event. This score is calibrated by comparing it with scores obtained from events generated by the background model at similar dijet masses. A high calibrated score therefore indicates that the event has a higher anomaly score than most of these background events.

The sensitivity of \RIDDLE{} is assessed against idealized anomaly detection (IAD) and Supervised, following the benchmarking strategy of Refs.~\cite{hallin2022cathode,hallin2023lacathode}, and against LaCATHODE~\cite{hallin2023lacathode} as an independent anomaly-detection method. IAD and Supervised share the Stein-witness construction and event-scoring procedure of \RIDDLE{}. In IAD, the witness is still trained on the unlabeled signal-region mixture, but its background reference is fitted to an independent sample of pure-background events from the signal region rather than inferred from sidebands. IAD therefore assesses the effect of sideband-based background estimation without supplying signal labels. In Supervised, the witness is instead trained on known-signal events, while a separate known-background sample is used to fit the reference. Supervised therefore shows how well the method separates signal from background when the signal is known during training. 

The method is evaluated using simulated dijet events from the LHC Olympics dataset~\cite{kasieczka2021lhco}. The analysis begins with four auxiliary observables: the two jet-mass variables, $m_1$ and $\Delta m$, and the two $N$-subjettiness ratios, $\tau_{21}^{J_1}$ and $\tau_{21}^{J_2}$. Two further tests, motivated by Ref.~\cite{hallin2023lacathode}, examine how the method responds to changes in these inputs. The first adds the angular separation between the two jets, $\Delta R$, as a fifth observable. The second shifts the two jet-mass observables to introduce an explicit dependence on $\mjj$, while retaining the original four-dimensional feature space. For each feature configuration, signal--background discrimination is evaluated, projections of the learned witness are examined, and background mass sculpting under anomaly-score selection is assessed using background-only signal-region events. Finally, the sensitivity of \RIDDLE{} is evaluated as a function of the signal fraction in the unlabeled training sample.
\section{\RIDDLE{}: conditional Stein-witness construction}\label{sec:method}

\subsection{A sideband-derived background reference}

Let $m=\mjj$ denote the dijet mass, $\xb\in\mathbb R^d$ the auxiliary event observables, and $\zb=T_\phi(\xb;m)$ their learned latent representation. The mapping $T_\phi$ is fitted using events in mass sidebands (SB), outside the signal region (SR). Its dependence on $m$ allows changes in QCD jet observables across the dijet-mass spectrum to be described, following earlier sideband and latent-space approaches~\cite{nachman2020anode,hallin2022cathode,hallin2023lacathode}. With this mapping held fixed, a conditional density model is fitted to a separate sample of mapped sideband events. This additional fit accounts for residual departures from the Gaussian latent distribution targeted by the mapping and constitutes the background correction. The resulting differentiable density $q(\zb\mid m)$ describes the latent background distribution at a given dijet mass, rather than averaging over mass to obtain $q(\zb)$.

The background reference enters witness training through its \emph{density score}, defined as
\begin{equation}
\bm s_q(\zb,m)=\nabla_{\zb}\log q(\zb\mid m).
\label{eq:densityscore}
\end{equation}
At fixed dijet mass, each component of this vector gives the derivative of the background log density with respect to one latent coordinate~\cite{hyvarinen2005score}. This vector supplies the background-dependent term in the Stein operator. Applying the operator to the witness function gives the \emph{Stein response} for each event. 

For witness training, this response is evaluated on unlabeled SR events transformed by the same sideband-trained mapping. Their latent coordinates and dijet masses define the joint distribution $p(\zb,m)$ used for training. The witness is optimized on this population to identify departures from the conditional background reference, while both the mapping and reference density remain fixed.

\subsection{Learning the scalar Stein witness}

A Stein test uses a vector-valued function $\bm f(\zb)$ to probe whether the observed events follow a reference distribution. For a differentiable $\bm f$, the Langevin--Stein operator~\cite{gorham2015stein} associated with the conditional background density $q(\zb\mid m)$ is denoted by $\mathcal T_q$ and defined as
\begin{equation}
\mathcal T_q\bm f
=\nabla_{\zb}\!\cdot\bm f
+\bm f\cdot\nabla_{\zb}\log q(\zb\mid m).
\label{eq:vectorstein}
\end{equation}
The operator depends on the reference through its density score~\cite{grathwohl2020stein,repasky2023critic}. Stein's identity states that the mean operator response under the reference distribution is zero for suitable test functions and boundary conditions~\cite{stein1972bound,grathwohl2020stein}. Learned Stein critics optimize the test function so that its response on the observed data can reveal departures from this reference expectation~\cite{grathwohl2020stein,repasky2023critic}.

In \RIDDLE{}, the witness $u_\theta(\zb)$ is a neural network with a single scalar output. Rather than learning an unrestricted vector field, the test field is defined by the gradient of this scalar function, $\bm f=\nabla_{\zb}u_\theta(\zb)$. Applying $\mathcal T_q$ to this gradient gives the scalar Stein operator $\mathcal A_q$, defined by $\mathcal A_q u_\theta=\mathcal T_q(\nabla_{\zb}u_\theta)$. Using Eq.~\eqref{eq:vectorstein}, its action on the witness becomes
\begin{equation}
\mathcal A_q u_\theta(\zb,m)
=\Delta_{\zb}u_\theta(\zb)
+\nabla_{\zb}u_\theta(\zb)\cdot\bm s_q(\zb,m).
\label{eq:stein}
\end{equation}
Both derivatives act only on the latent coordinates. The Laplacian sums the second derivative of $u_\theta$ with respect to each of the $d$ latent coordinates. The background density score is evaluated at each event's dijet mass, while the witness network receives only $\zb$, with no direct mass input.

At a fixed dijet mass, the mean Stein response is obtained by averaging Eq.~\eqref{eq:stein} over the conditional background density $q(\zb\mid m)$. This average is denoted by $\mathbb E_{q(\zb\mid m)}$. The first term, $\Delta_{\zb}u$, describes the local curvature of the witness. The second, $\nabla_{\zb}u\cdot\bm s_q$, describes how the witness changes in the direction of increasing background density. Under the conditions stated in Appendix~\ref{app:stein}, integration by parts shows that these two terms make equal and opposite contributions to the background average. Their cancellation gives
\begin{equation}
\mathbb E_{q(\zb\mid m)}[\mathcal A_q u(\zb,m)]=0.
\label{eq:steinidentity}
\end{equation}
For an unbounded latent space, the boundary contribution vanishes if the density-weighted witness gradient, $q(\zb\mid m)\nabla_{\zb}u(\zb)$, decays sufficiently rapidly at large distances. The explicit condition and derivation are given in Appendix~\ref{app:stein}. The identity concerns an average over the reference distribution; neither an individual event's response nor an average over a finite sample must be exactly zero. If the SR distribution $p(\zb\mid m)$ differs from $q(\zb\mid m)$, the expected response can instead be nonzero. The witness is trained to reveal such differences without fitting a separate signal distribution.

A single witness network is trained on events throughout the SR mass interval, rather than fitting a separate network at each mass. Both expectations in the loss are taken over the joint SR distribution $p(\zb,m)$, with $\mathbb E_p$ denoting an average over this distribution. These expectations are estimated by averaging over training events, so mass intervals containing more events contribute more to the loss. The principal regularized loss is
\begin{equation}
\mathcal L_{\mathrm{Stein}}(\theta)
=-\mathbb E_p[\mathcal A_q u_\theta]
+\frac{\lambda}{2}\mathbb E_p[\|\nabla_{\zb}u_\theta\|^2].
\label{eq:objective}
\end{equation}
Minimizing the first term increases the mean Stein response on the SR data. The second term penalizes large witness gradients, with $\lambda$ controlling the penalty strength. This is an $L^2$ regularization of the test field $\bm f=\nabla_{\zb}u_\theta$, following the motivation of regularized neural Stein critics~\cite{grathwohl2020stein,repasky2023critic}. Here, the regularization strength is held fixed at $\lambda=1$ throughout training.

The Stein loss constrains the average response over the training population but does not directly optimize the ranking of high-scoring events. An auxiliary \emph{tail-ranking term} therefore focuses on these events, which are particularly important for strong background rejection. Events are ranked using $P=-u_\theta$, so lower witness values correspond to higher scores. High-scoring SR and generated-background events are selected, and the loss penalizes cases in which a selected background event scores higher than a selected SR event. This encourages better separation between the selected SR events and background events that are difficult to reject. The SR events are selected using their witness scores, without knowledge of their true signal or background labels. The full training loss is $\mathcal L_{\mathrm{Stein}}+\eta\mathcal L_{\mathrm{tail}}$, where $\eta$ controls the contribution of the auxiliary term. The loss definition and numerical settings are given in Appendix~\ref{app:numerical}.

\subsection{From the witness to an event score}

The witness is trained to reveal differences between the SR distribution and the background reference, but an anomaly search also requires a score for each event. \RIDDLE{} constructs this score from three quantities derived from the witness: the negative witness output, $P=-u_\theta$; the squared gradient magnitude, $E=\|\nabla_{\zb}u_\theta\|^2$; and the Stein response, $W=\mathcal A_q u_\theta$. Each quantity is separately converted to a standard-normal value using its percentile among generated-background events at similar dijet masses. The normalized quantities are then combined into a score, giving additional weight to unusually high Stein responses. The scores from the retained witness networks are averaged within each fit and then across the repeated fits to obtain $R_{\rm ensemble}$.

A support correction, $\Pi(\zb,m)$, is then applied to reduce scores for events with unusually large latent distances or background log-density gradients. Such events may be more sensitive to errors in the background model. The corrected scores are then calibrated by comparing them with the corrected scores of a separate generated-background sample at similar dijet masses. Let $F_q(r\mid m)$ denote the fraction of reference background events with corrected scores at or below $r$ at comparable dijet mass. The final event score is
\begin{equation}
S(\zb,m)=
\bigl[
F_q(R_{\rm ensemble}(\zb,m)-\Pi(\zb,m)\mid m)
\bigr]^{10}.
\label{eq:score}
\end{equation}
An event receives a high final score when its corrected score exceeds those of most reference background events at similar dijet masses. This indicates how unusual the event is relative to the background, but the score is neither an exact likelihood ratio nor a signal probability. Details of the normalization, ensemble averaging, support correction, and calibration are given in Appendix~\ref{app:numerical}.

The mapping, background reference, witness ensemble, and calibration are fixed before the held-out events are evaluated. The same scoring procedure is used in the SR and sidebands, without training a separate witness for sideband events. For events outside the fitted mass interval, the nearest supported boundary mass is used as the context for conditional scoring.

\subsection{IAD and Supervised: information-access benchmarks}\label{subsec:benchmarks}

IAD and Supervised assess the sensitivity of \RIDDLE{} when known-background or known-signal events are available for training. Both use the same background-mapping procedure, scalar Stein-witness construction, event-scoring rules, checkpoint selection, ensemble aggregation, support correction, and background-based calibration as \RIDDLE{}. Neither uses a separate binary classifier.

In \RIDDLE{}, the witness is trained on the unlabeled SR mixture and the background reference is inferred from sidebands. IAD uses the same unlabeled mixture for witness training but fits the reference to an independent pure-QCD sample from the SR. It therefore tests the sensitivity achievable when representative background events are available directly, allowing the effect of sideband-based background estimation to be assessed without supplying signal labels to the witness.

In Supervised, the witness is instead trained on a separate sample of known-signal events, while the background reference is fitted using a separate sample of known-QCD events. This benchmark shows how well the same method separates signal from background when examples of the target signal are available during training. The reference, witness, and calibration are fitted using the samples specified for each configuration; the shared procedure does not require identical fitted models. IAD and Supervised provide performance references, rather than guaranteed upper bounds on the sensitivity of \RIDDLE{}.

In the background-only scenario, no signal is injected into the ordinary SR mixture. Supervised nevertheless retains its independent known-signal training sample; the background-only designation does not remove the signal information available to this benchmark.
\section{Dijet simulation and evaluation strategy}\label{sec:data}

\subsection{Signal, background, and physical observables}\label{subsec:features}
The numerical study uses the two-prong LHC Olympics R\&D dataset~\cite{kasieczka2019lhcord}, supplemented by its additional QCD sample~\cite{shih2021additionalqcd}. The background consists of simulated QCD dijet events. The benchmark signal follows $W'\to XY\to(q\bar q)(q\bar q)$, with $m_{W'}=3.5\TeV$, $m_X=500\,\GeV$, and $m_Y=100\,\GeV$. The dijet invariant mass, $m\equiv\mjj$, is used to search for a localized signal excess over the QCD background. The signal region (SR) is defined by $3.3\TeV<\mjj<3.7\TeV$. Events with dijet masses outside this interval, but within the prepared mass range, form the sidebands used to model the background.

The two highest-$p_{\mathrm T}$ jets in each event are used to form the dijet system. These jets are reconstructed in the original LHCO dataset using the anti-$k_T$ algorithm~\cite{cacciari2008antikt} with radius parameter $R=1$, and no additional jet reconstruction or jet-level cuts are applied in this study. Following Ref.~\cite{hallin2022cathode}, the two jets are ordered by increasing jet mass, $m_{J_1}\leq m_{J_2}$, regardless of their transverse-momentum ordering. The auxiliary observables are
\begin{equation}
\xb=\bigl(m_1,\,\Delta m,\,\tau_{21}^{J_1},\,\tau_{21}^{J_2}\bigr),
\qquad d=4,
\label{eq:features}
\end{equation}
where $m_1\equiv m_{J_1}$ is the lower of the two jet masses and $\Delta m\equiv m_{J_2}-m_{J_1}\geq0$ is their mass difference. The two $N$-subjettiness ratios describe the substructure of the lower- and higher-mass jets, respectively, and help distinguish two-prong from one-prong jet structure~\cite{thaler2011nsubjettiness}.

The dijet mass is not included in $\xb$. Instead, it conditions the background mapping and density model, as well as the final score calibration described in Sec.~\ref{sec:method}. The transformed coordinates $\zb=T_\phi(\xb;m)$ are the learned latent representation of these observables, not additional physical inputs.

\subsection{Statistical populations}\label{subsec:populations}

The simulated event samples are divided into nonoverlapping subsets for training, validation, and held-out evaluation. Table~\ref{tab:populations} summarizes the training and validation event counts for the signal-injected analyses. Some subsets are shared by \RIDDLE{}, IAD, and Supervised; these contain the same events rather than additional independent samples. The same event assignments are retained for the $\Delta R$ and shifted-feature studies described in Secs.~\ref{sec:deltar} and~\ref{sec:shifted}. The roles of the different training populations for the three methods are explained in Sec.~\ref{subsec:benchmarks}.

The held-out SR evaluation sample contains 386,472 QCD events and 25,065 signal events and is separate from the SR data used for training. The background-only evaluation uses the same QCD events without injected signal. In the signal-injection study, the number of injected signal events is varied while the evaluation sample remains unchanged. The background-only training and validation populations are described in Appendix~\ref{app:bgpopulations}.

Simulation truth is used to prepare the pure-QCD reference samples for IAD and Supervised and the known-signal training sample for Supervised, as well as to evaluate performance. In \RIDDLE{}, model training, validation, selection, and anomaly-score construction and calibration do not use signal or background truth labels.

LaCATHODE~\cite{hallin2023lacathode} is evaluated using the authors' public implementation, with the same prepared datasets, auxiliary observables, and held-out evaluation populations. The method learns a classifier in the background-trained latent representation. Independent background-flow and classifier fits are used for each random seed, retaining the original validation-based classifier checkpoint-selection procedure. All methods are evaluated at the same background acceptances.

\begin{table}[H]
\centering
\caption{Training and validation event counts for signal-injected Stein-witness analyses.}
\label{tab:populations}

\begin{tabular*}{\linewidth}
  {@{}l@{\extracolsep{\fill}}rr@{}}
\hline\hline
Population & Training & Validation \\
\hline

\parbox[t]{0.63\linewidth}{\raggedright
SB map (\RIDDLE{}, IAD, Supervised)}
& 307,387 & 73,436 \\

\parbox[t]{0.63\linewidth}{\raggedright
SB correction (\RIDDLE{})}
& 131,738 & 36,718 \\

\parbox[t]{0.63\linewidth}{\raggedright
SR mixture (\RIDDLE{}, IAD)}
& 49,100 & 12,275 \\

\parbox[t]{0.63\linewidth}{\raggedright
Pure SR QCD (IAD, Supervised)}
& 199,791 & 26,720 \\

\parbox[t]{0.63\linewidth}{\raggedright
Known SR signal (Supervised)}
& 29,810 & 7,453 \\[2pt]

\hline\hline
\end{tabular*}
\end{table}

\subsection{Discrimination measures}\label{subsec:metrics}

Signal--background discrimination is evaluated using receiver operating characteristic (ROC) curves and the area under the ROC curve (AUC). At each anomaly-score threshold $t$, let $\epsS(t)$ and $\epsB(t)$ denote the fractions of signal and QCD events passing the selection. Varying this threshold gives the ROC curve, which shows the signal efficiency as a function of the background efficiency, while the AUC summarizes the overall discrimination performance.

To assess the improvement in sensitivity from event selection, the significance-improvement characteristic (SIC) is used~\cite{black2011sic}:
\begin{equation}
\SIC(t)=\frac{\epsS(t)}{\sqrt{\epsB(t)}}.
\label{eq:sic}
\end{equation}
The SIC measures the improvement in the approximate counting sensitivity $S/\sqrt B$ relative to the unselected sample. Its maximum value provides a measure of discrimination that emphasizes signal retention at low background efficiency. For both ROC and SIC curves, efficiencies are calculated using the full held-out SR evaluation sample.

For the signal-injection study described in Sec.~\ref{sec:injection}, the counting sensitivity after selection is estimated following Ref.~\cite{hallin2022cathode} as $Z_{\mathrm{count}}(t)=\SIC(t)\,N_{S,\SR}/\sqrt{N_{B,\SR}}$. Here, $N_{S,\SR}$ and $N_{B,\SR}$ are the signal and QCD event counts in the SR before selection at each injection strength, while the selection efficiencies are measured on the separate held-out evaluation sample. The SIC scales the uncut $S/\sqrt B$ to account for signal retention and background rejection after selection. This estimate uses the $S/\sqrt B$ approximation for the median counting significance, which is valid when the signal is small compared with the background and the background uncertainty is negligible~\cite{cowan2011asymptotic}.

\subsection{Background mass-shape stability}\label{subsec:massshape}

For a resonance search, anomaly-score selection must preserve the smoothly falling QCD dijet-mass spectrum. To ensure consistency across the mass spectrum, the same trained scoring method and thresholds are applied in the SR and sidebands. However, correlations between the anomaly score and $\mjj$ may still distort the background distribution.

Following Ref.~\cite{hallin2023lacathode}, these distortions are assessed using a $\chi^2/n_{\rm dof}$ test on the held-out background-only SR sample. The preselection QCD dijet-mass distribution is divided into 300 bins containing approximately equal event counts. At each background efficiency, the selected distribution is compared with the preselection distribution scaled to the selected event count. The shape difference is quantified by $\chi^2/n_{\rm dof}$, with $n_{\rm dof}=n_{\rm bins}-1$.

For comparison, the same $\chi^2/n_{\rm dof}$ calculation is repeated for 100 randomly selected QCD subsets at each background efficiency. The resulting statistical fluctuations provide a reference for assessing background mass-shape stability under anomaly-score selection.

\section{\RIDDLE{} results}\label{sec:mainresults}

The results in this section use the four auxiliary observables defined in Eq.~\eqref{eq:features}. The performance of \RIDDLE{} is compared with IAD and Supervised, which share the same Stein-witness construction but differ in the information available during training (Sec.~\ref{subsec:benchmarks}), and with LaCATHODE~\cite{hallin2023lacathode} as an independent anomaly-detection method. The analysis examines signal--background discrimination, the learned witness response, and the effect of anomaly-score selection on the dijet-mass spectrum.

\begin{figure*}[!t]
\centering
\plotpanel{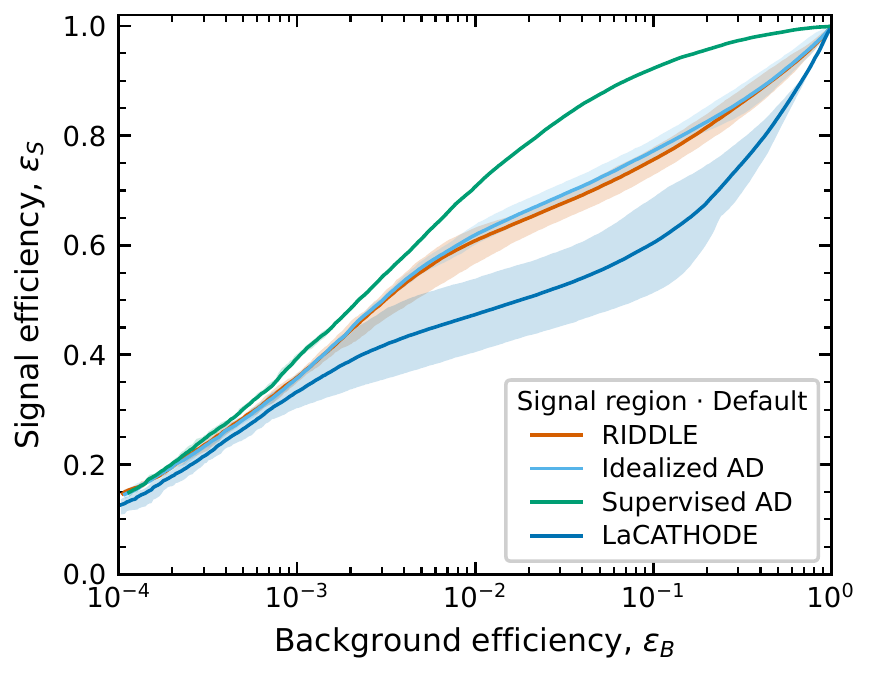}{}\hfill%
\plotpanel{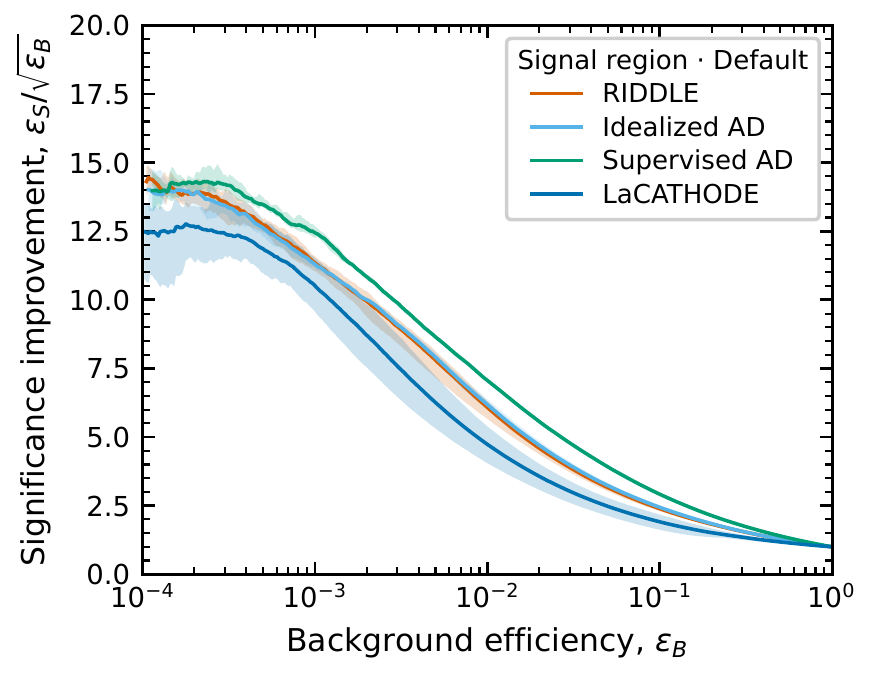}{}
\caption{Signal--background discrimination in the held-out SR using the original four observables. ROC (left) and SIC (right) compare \RIDDLE{}, IAD, Supervised, and LaCATHODE. Curves show means over five runs with different random seeds, with shaded bands indicating the minimum and maximum values across runs.}
\label{fig:main-discrimination}
\end{figure*}

Figure~\ref{fig:main-discrimination} compares the ROC and SIC curves, while Table~\ref{tab:perf-main} summarizes the AUC, maximum SIC, and signal efficiencies at fixed background acceptances. At the reference signal injection, \RIDDLE{} and IAD show comparable discrimination despite using different background references. While IAD uses independent pure-QCD events from the SR, \RIDDLE{} estimates the background distribution from sidebands. Their similar performance suggests that the sideband-derived reference provides sufficient information for effective signal--background separation in this simulated sample. However, this agreement does not imply that the fitted background distributions are identical.

Supervised achieves a higher AUC and retains more signal at the reported background acceptances, but its maximum SIC remains comparable to those of \RIDDLE{} and IAD. The improvement in overall discrimination therefore does not translate into a similarly large gain at the threshold that maximizes SIC. This distinction illustrates why AUC, maximum SIC, and signal efficiency at fixed background acceptance provide complementary measures of discrimination. 

Compared with LaCATHODE, \RIDDLE{} achieves higher mean AUC and maximum SIC and a relative increase of 25--29\% in mean signal efficiency at the reported background acceptances. This improvement is obtained without using signal labels during training.

\begin{table*}[!htbp]
\centering
\caption{Discrimination performance with the original four observables. Values are means over five runs; superscript and subscript deviations indicate the differences to the maximum and minimum values across runs, respectively.}
\label{tab:perf-main}
\renewcommand{\arraystretch}{1.5}
\begin{tabular*}{\linewidth}
{@{\extracolsep{\fill}}lcccccc@{}}
\hline\hline
 & \multicolumn{2}{c}{Discrimination}
 & \multicolumn{4}{c}{Signal efficiency at $\epsB$} \\
\cline{2-3}\cline{4-7}
Configuration & AUC & Max. SIC
 & $10\%$ & $5\%$ & $1\%$ & $0.5\%$ \\
\hline
\RIDDLE{} &
$0.885^{+0.02}_{-0.02}$ &
$14.56^{+0.40}_{-0.36}$ &
$0.756^{+0.02}_{-0.03}$ &
$0.706^{+0.02}_{-0.03}$ &
$0.608^{+0.02}_{-0.04}$ &
$0.552^{+0.02}_{-0.04}$ \\
IAD &
$0.890^{+0.02}_{-0.01}$ &
$14.45^{+0.47}_{-0.42}$ &
$0.772^{+0.02}_{-0.02}$ &
$0.723^{+0.02}_{-0.02}$ &
$0.619^{+0.02}_{-0.02}$ &
$0.559^{+0.01}_{-0.01}$ \\
Supervised &
$0.968^{+0.00}_{-0.00}$ &
$14.61^{+0.43}_{-0.40}$ &
$0.922^{+0.00}_{-0.00}$ &
$0.875^{+0.00}_{-0.00}$ &
$0.707^{+0.00}_{-0.00}$ &
$0.612^{+0.00}_{-0.00}$ \\
LaCATHODE &
$0.807^{+0.03}_{-0.02}$ &
$13.04^{+0.68}_{-1.40}$ &
$0.604^{+0.08}_{-0.09}$ &
$0.554^{+0.07}_{-0.08}$ &
$0.473^{+0.07}_{-0.07}$ &
$0.443^{+0.07}_{-0.07}$ \\[2pt]
\hline\hline
\end{tabular*}
\end{table*}

\begin{figure*}[!t]
\centering
\plotpanel{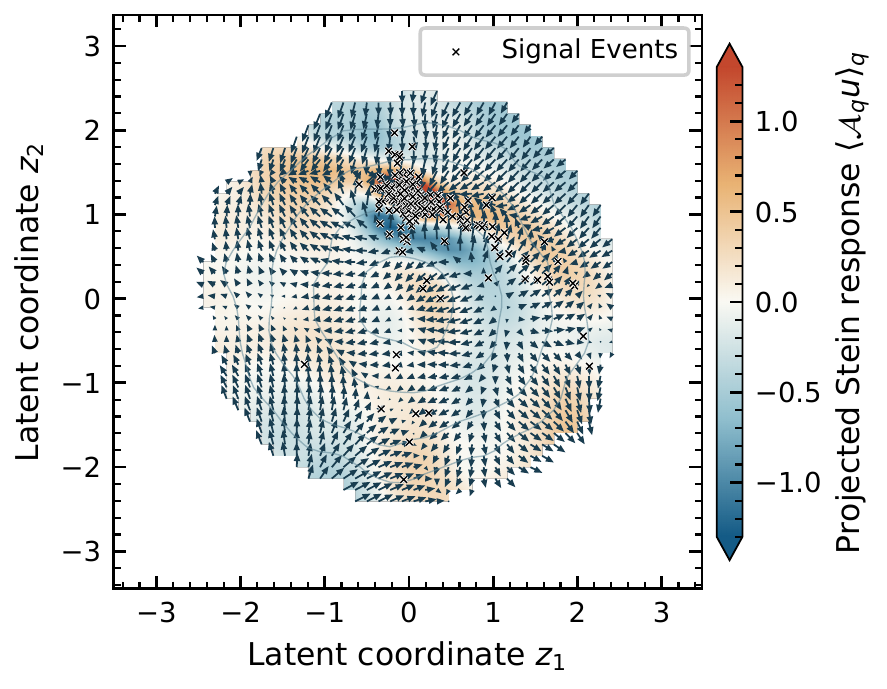}{(a) Signal-injected SR}\hfill%
\plotpanel{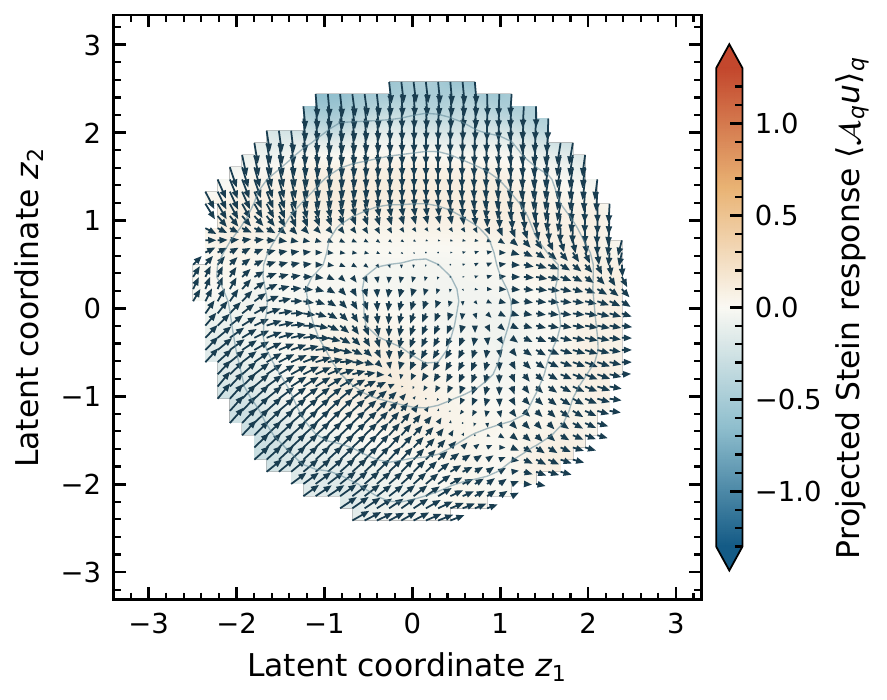}{(b) Background-only SR}
\caption{Projected \RIDDLE{} Stein responses for the original four observables, using witnesses trained separately on (a) signal-injected and (b) background-only SR samples. Colors show binned, smoothed averages of $\mathcal A_q u_\theta$ over generated-background reference events and the witness ensemble; arrows show corresponding averages of $-\nabla_{\zb}u_\theta$, indicating directions of decreasing witness values. Both are evaluated in four-dimensional latent space before projection onto $(z_1,z_2)$.}
\label{fig:main-witness}
\end{figure*}

The learned witness response is examined in Figure~\ref{fig:main-witness}, which compares two \RIDDLE{} witnesses trained separately on the signal-injected and background-only SR samples. In the signal-injected case, localized structures in the projected Stein response overlap with regions populated by simulated signal events. The signal-event positions are overlaid using simulation truth only after training and are not used to learn the witness. The background-only case shows significantly less pronounced variations in the projected response. These projections illustrate the learned witness response but do not represent the final calibrated anomaly score.

\begin{figure*}[!t]
\centering
\plotpanel{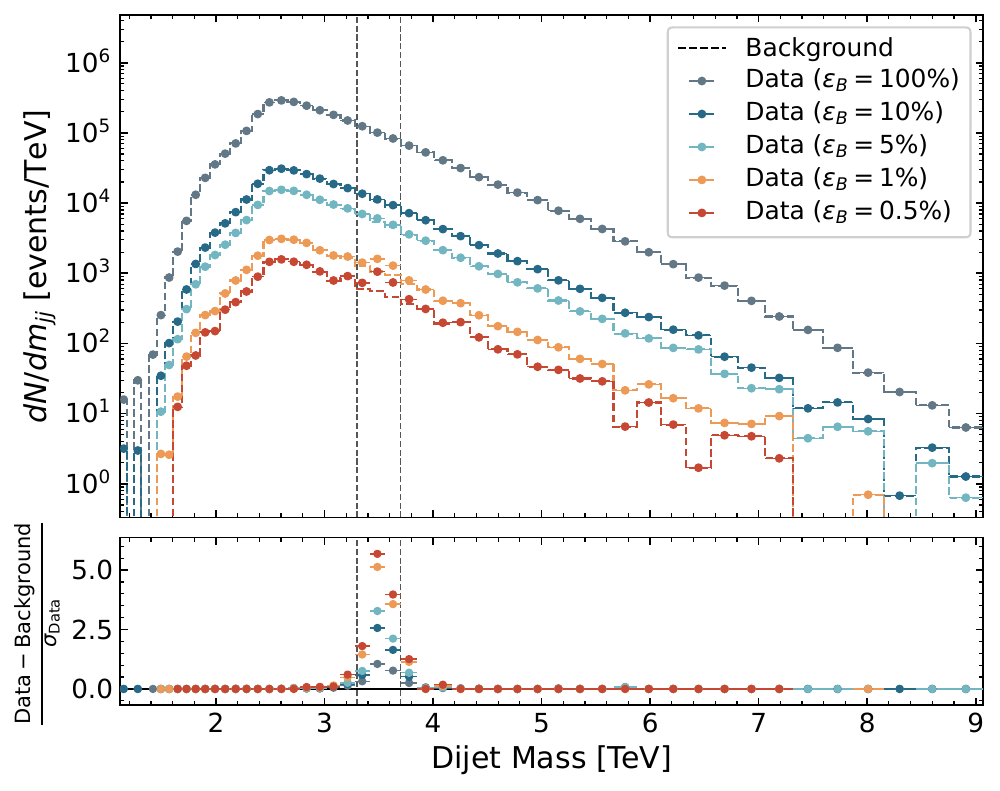}{}\hfill%
\plotpanel{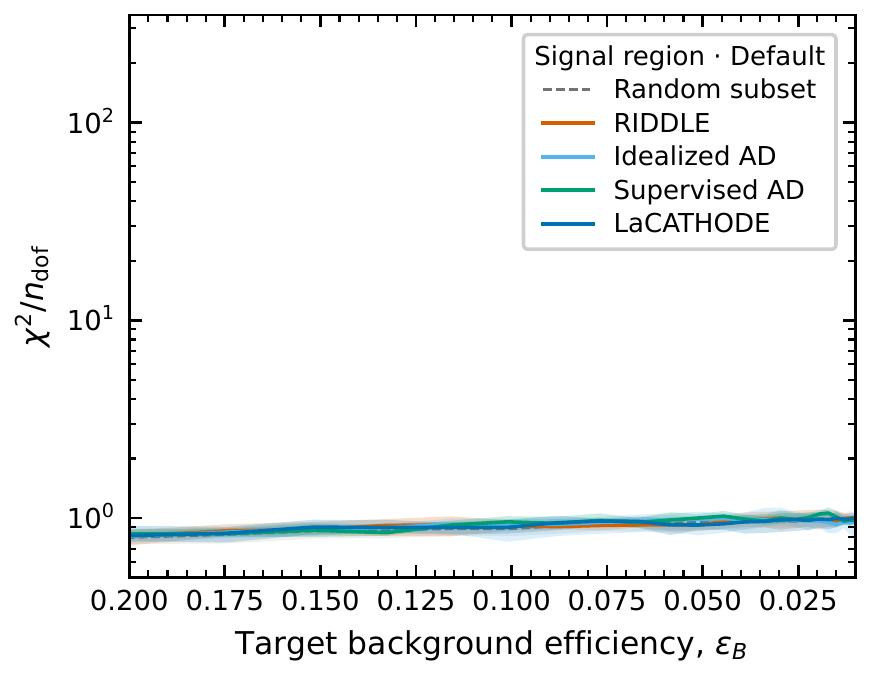}{}
\caption{Dijet-mass spectrum and background mass-sculpting test using the original four observables. Left: signal-injected \RIDDLE{} spectrum before and after selection at the indicated background acceptances, using the same score thresholds in the SR and sidebands. Right: background-only SR $\chi^2/n_{\rm dof}$ for \RIDDLE{}, IAD, Supervised, and LaCATHODE. Colored curves show means over five runs, with shaded bands spanning the run minima and maxima. The gray dashed line shows the median of 100 random QCD selections at matching background efficiencies.}
\label{fig:mass-main}
\end{figure*}

Figure~\ref{fig:mass-main} shows the dijet-mass spectrum before and after anomaly-score selection, together with a separate background mass-sculpting test. As the \RIDDLE{} score requirement becomes more restrictive, the injected signal becomes more prominent relative to the QCD background. The same scoring procedure and selection thresholds are applied in both the SR and sidebands to examine the background mass distribution across the dijet-mass range.

The background-only $\chi^2/n_{\rm dof}$ comparison provides a quantitative test of mass-shape stability, as described in Sec.~\ref{subsec:massshape}. All four methods remain close to the random-selection reference across the tested background efficiencies, indicating that the QCD dijet-mass distribution remains largely stable under anomaly-score selection in the SR. The higher signal retention of \RIDDLE{} relative to LaCATHODE is thus achieved with comparable background mass-shape stability. Together with the more pronounced signal peak in the SR, these results demonstrate the ability of \RIDDLE{} to suppress the QCD background while retaining a substantial fraction of signal events.

\resultsectionbreak
\section{Adding interjet angular separation \texorpdfstring{$\Delta R$}{Delta R}}\label{sec:deltar}

The interjet angular separation $\Delta R$ provides additional information about the event geometry but is also correlated with the dijet mass. Including it therefore tests both signal--background discrimination and the preservation of the background mass shape under anomaly-score selection. Following Ref.~\cite{raine2023curtains,hallin2023lacathode}, $\Delta R$ is added as a fifth auxiliary observable alongside the original four inputs. The interjet separation is defined as
\begin{equation}
\Delta R=\sqrt{(\Delta\eta)^2+(\Delta\varphi)^2},
\label{eq:deltar}
\end{equation}
where $\Delta\eta=\eta_{J_2}-\eta_{J_1}$ is the difference in jet pseudorapidity and $\Delta\varphi=\varphi_{J_2}-\varphi_{J_1}$ is the azimuthal-angle difference, taken as the shortest angular separation between the jets. The input vector is therefore
\[
\xb_{\Delta R}=\bigl(m_1,\,\Delta m,\,\tau_{21}^{J_1},\,\tau_{21}^{J_2},\,\Delta R\bigr),
\]
which is mapped to a five-dimensional latent representation. The signal model and SR definition are unchanged, and the training, validation, and evaluation samples contain the same events as in the four-observable analysis. All four methods are retrained with these five inputs using their respective training and scoring procedures.

\begin{figure*}[!t]
\centering
\plotpanel{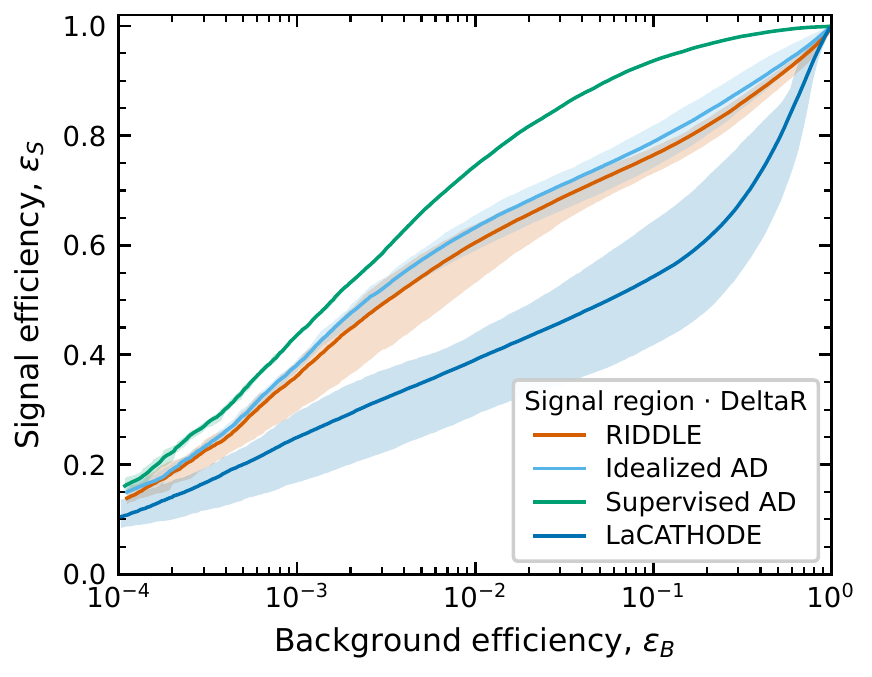}{}\hfill%
\plotpanel{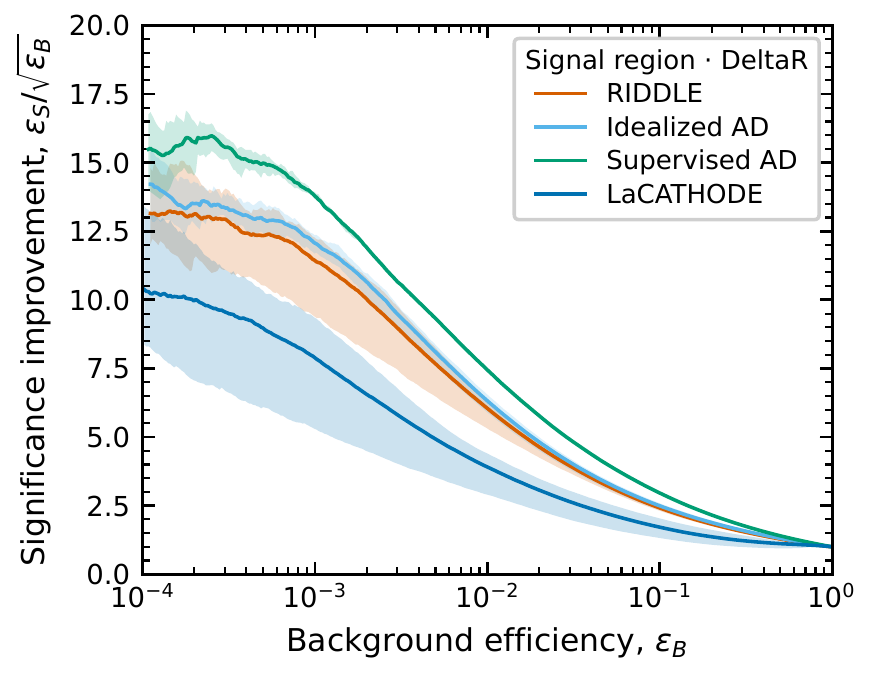}{}
\caption{Signal--background discrimination in the held-out SR with interjet $\Delta R$ added to the original four observables. ROC (left) and SIC (right) compare \RIDDLE{}, IAD, Supervised, and LaCATHODE. Curves show means over five runs with different random seeds, with shaded bands indicating the minimum and maximum values across runs.}
\label{fig:deltar-discrimination}
\end{figure*}

Figure~\ref{fig:deltar-discrimination} and Table~\ref{tab:perf-deltaR} summarize the discrimination performance. For \RIDDLE{}, the AUC and signal efficiencies remain close to those with the original four observables, while the mean maximum SIC is slightly lower with wider variation across runs. IAD retains a similar maximum SIC, whereas Supervised improves across all reported metrics, benefiting more clearly from the additional angular information. Compared with LaCATHODE, \RIDDLE{} achieves a relative increase of 41--56\% in mean signal efficiency at the reported background acceptances and approximately 27\% in mean maximum SIC.

\begin{table*}[!htbp]
\caption{Discrimination performance with interjet $\Delta R$ added as a fifth observable. Values are means over five runs; superscript and subscript deviations indicate the differences to the maximum and minimum values across runs, respectively.}
\label{tab:perf-deltaR}
\renewcommand{\arraystretch}{1.5}
\begin{tabular*}{\linewidth}
{@{\extracolsep{\fill}}lcccccc@{}}
\hline\hline
 & \multicolumn{2}{c}{Discrimination} & \multicolumn{4}{c}{Signal efficiency at $\epsB$} \\
\cline{2-3}\cline{4-7}
Configuration & AUC & Max. SIC & $10\%$ & $5\%$ & $1\%$ & $0.5\%$ \\
\hline
\RIDDLE{} &
$0.886^{+0.01}_{-0.02}$ &
$13.64^{+1.46}_{-1.40}$ &
$0.764^{+0.01}_{-0.03}$ &
$0.719^{+0.02}_{-0.04}$ &
$0.605^{+0.03}_{-0.07}$ &
$0.542^{+0.04}_{-0.08}$ \\
IAD &
$0.903^{+0.02}_{-0.02}$ &
$14.51^{+1.52}_{-1.31}$ &
$0.788^{+0.03}_{-0.05}$ &
$0.741^{+0.03}_{-0.05}$ &
$0.632^{+0.02}_{-0.04}$ &
$0.573^{+0.02}_{-0.03}$ \\
Supervised &
$0.974^{+0.00}_{-0.00}$ &
$16.38^{+0.54}_{-0.81}$ &
$0.936^{+0.00}_{-0.00}$ &
$0.892^{+0.00}_{-0.00}$ &
$0.745^{+0.00}_{-0.00}$ &
$0.659^{+0.00}_{-0.00}$ \\
LaCATHODE &
$0.771^{+0.06}_{-0.09}$ &
$10.76^{+2.73}_{-2.36}$ &
$0.543^{+0.10}_{-0.12}$ &
$0.492^{+0.08}_{-0.12}$ &
$0.391^{+0.05}_{-0.10}$ &
$0.349^{+0.05}_{-0.10}$ \\[2pt]
\hline\hline
\end{tabular*}
\end{table*}

\begin{figure*}[!t]
\centering
\plotpanel{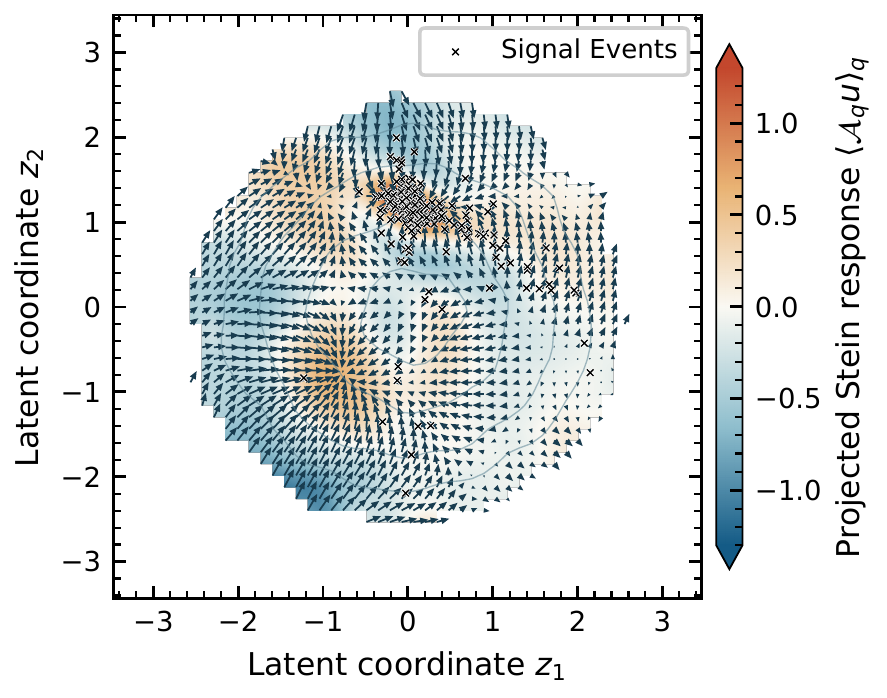}{(a) Signal-injected SR}\hfill%
\plotpanel{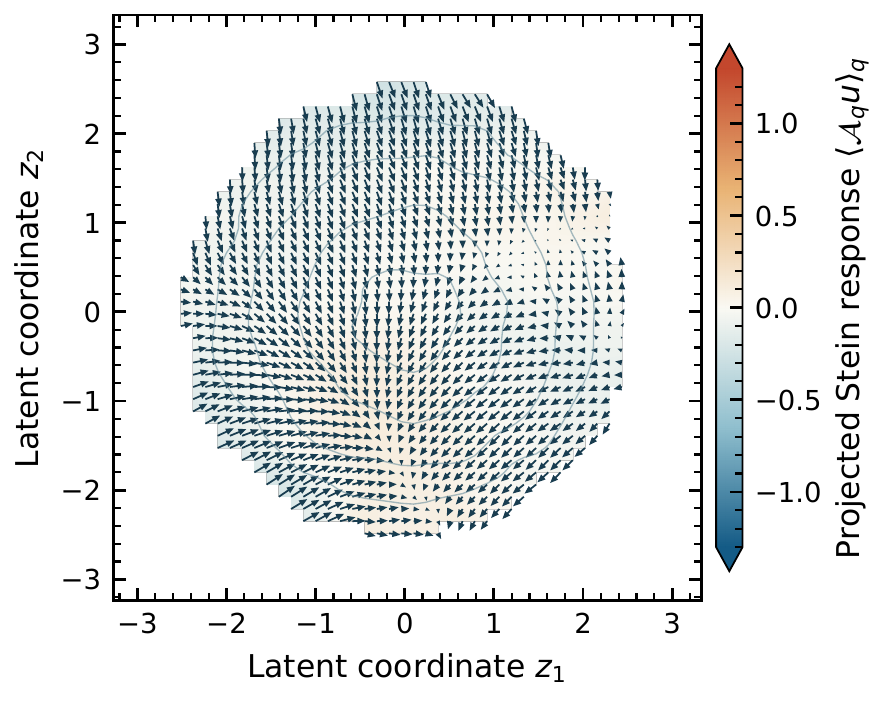}{(b) Background-only SR}
\caption{Projected \RIDDLE{} Stein responses with interjet $\Delta R$ added to the original four observables, using witnesses trained separately on (a) signal-injected and (b) background-only SR samples. Colors show binned, smoothed averages of $\mathcal A_q u_\theta$ over generated-background reference events and the witness ensemble; arrows show corresponding averages of $-\nabla_{\zb}u_\theta$, indicating directions of decreasing witness values. Both are evaluated in five-dimensional latent space before projection onto $(z_1,z_2)$.}
\label{fig:deltar-witness}
\end{figure*}

Figure~\ref{fig:deltar-witness} compares two \RIDDLE{} witnesses trained separately on the signal-injected and background-only SR samples. As in the four-observable analysis, the signal-injected projection shows localized structure near the overlaid signal events, while the background-only response is less pronounced. The signal positions are added using simulation truth after training and are not used to learn the witness. Although only $(z_1,z_2)$ is displayed, the response and witness gradient are evaluated using all five latent coordinates. The figure illustrates the learned witness response rather than the final event score.

\begin{figure*}[!t]
\centering
\plotpanel{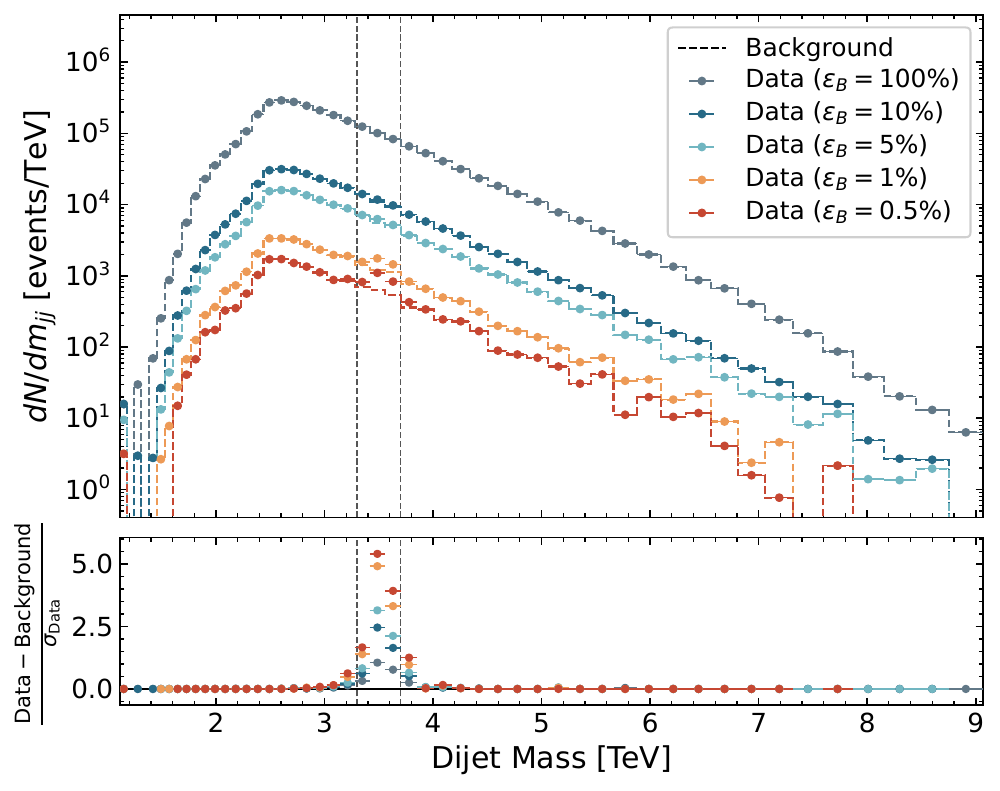}{}\hfill%
\plotpanel{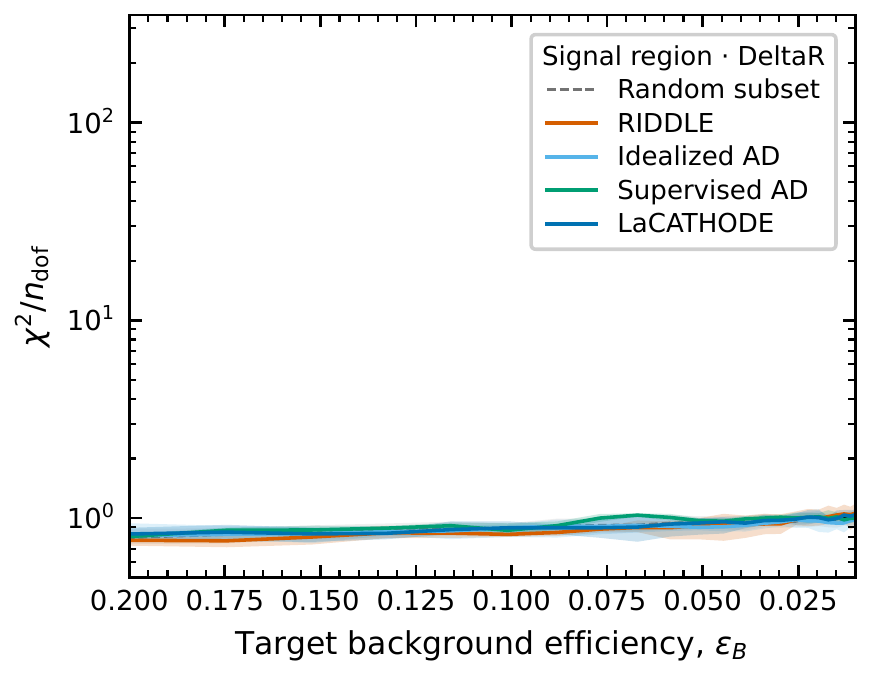}{}
\caption{Dijet-mass spectrum and background mass-sculpting test with interjet $\Delta R$ added to the original four observables. Left: signal-injected \RIDDLE{} spectrum before and after selection at the indicated background acceptances, using the same score thresholds in the SR and sidebands. Right: background-only SR $\chi^2/n_{\rm dof}$ for \RIDDLE{}, IAD, Supervised, and LaCATHODE. Colored curves show means over five runs, with shaded bands spanning the run minima and maxima. The gray dashed line shows the median of 100 random QCD selections at matching background efficiencies.}
\label{fig:mass-deltar}
\end{figure*}

Figure~\ref{fig:mass-deltar} shows the dijet-mass spectrum before and after anomaly-score selection, together with the background mass-sculpting test. As in the four-observable analysis, the same scoring procedure and thresholds are applied in the SR and sidebands. The spectrum shows that tighter score requirements suppress the QCD background relative to the signal peak in the SR. In the mass-sculpting test, all four methods yield background-only $\chi^2/n_{\rm dof}$ values close to the random-selection reference across the tested background efficiencies, indicating that the QCD mass shape within the SR remains largely stable when $\Delta R$ is included.

\resultsectionbreak
\section{Testing deliberately mass-shifted features}\label{sec:shifted}

The effect of mass correlations is further examined by shifting the jet-mass observables by an amount proportional to $\mjj$. Unlike the $\Delta R$ analysis in Sec.~\ref{sec:deltar}, this test modifies the original inputs without adding an observable, assessing whether discrimination and background mass-shape stability are maintained under explicit mass dependence. Following Ref.~\cite{hallin2023lacathode}, the two jet-mass observables are replaced by
\begin{equation}
m_1'=m_1+0.1\mjj,\qquad \Delta m'=\Delta m+0.1\mjj.
\label{eq:shift}
\end{equation}
The two $N$-subjettiness ratios are unchanged, so $\xb'=(m_1',\Delta m',\tau_{21}^{J_1},\tau_{21}^{J_2})$ remains four-dimensional. The signal model, SR definition, and event assignments remain unchanged from the original and $\Delta R$ analyses.  All four methods are retrained on the shifted inputs using their respective training and scoring procedures.

\begin{figure*}[!t]
\centering
\plotpanel{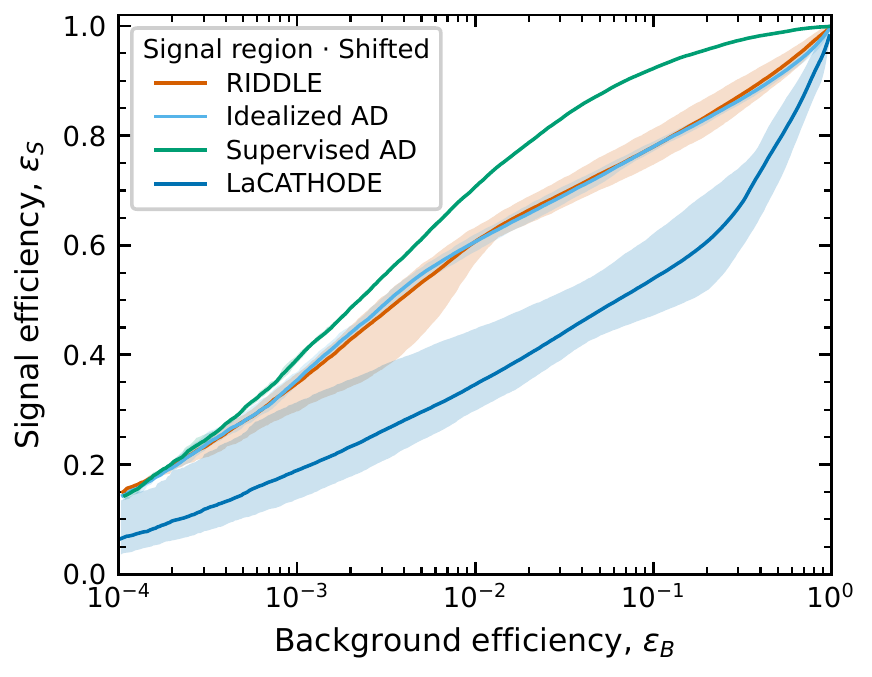}{}\hfill%
\plotpanel{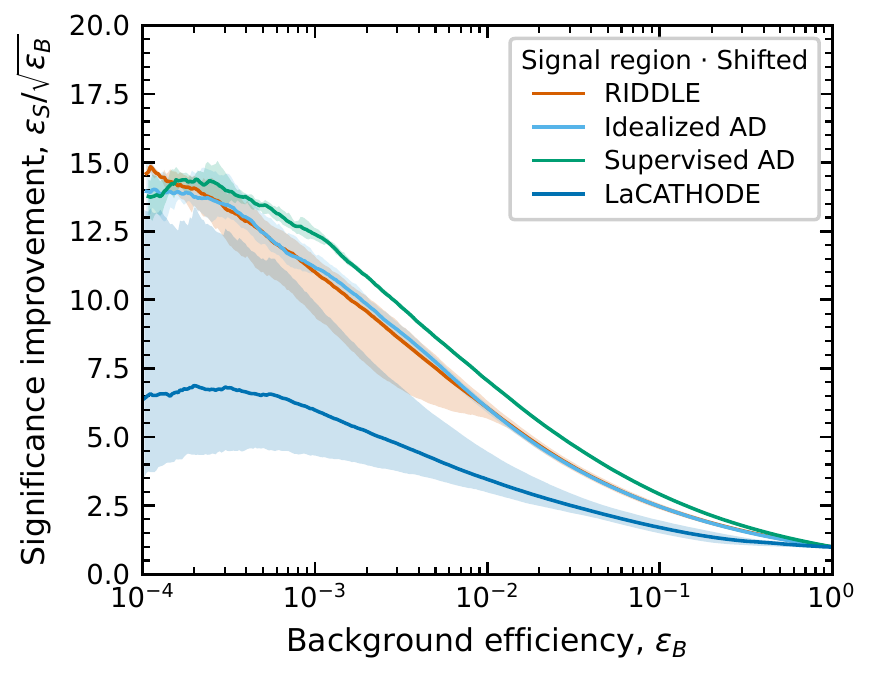}{}
\caption{Signal--background discrimination in the held-out SR using the mass-shifted observables of Eq.~\eqref{eq:shift}. ROC (left) and SIC (right) compare \RIDDLE{}, IAD, Supervised, and LaCATHODE. Curves show means over five runs with different random seeds, with shaded bands indicating the minimum and maximum values across runs.}
\label{fig:shift-discrimination}
\end{figure*}

Figure~\ref{fig:shift-discrimination} and Table~\ref{tab:perf-shifted} summarize the discrimination performance. \RIDDLE{}, IAD, and Supervised retain performance close to their four-observable results (Table~\ref{tab:perf-main}). In contrast to the reduced discrimination of LaCATHODE, \RIDDLE{} achieves a relative increase of 45--80\% in mean signal efficiency at the reported background acceptances and more than twice the mean maximum SIC of LaCATHODE.

\begin{table*}[!htbp]
\caption{Discrimination performance with the mass-shifted observables. Values are means over five runs; superscript and subscript deviations indicate the differences to the maximum and minimum values across runs, respectively.}
\label{tab:perf-shifted}
\renewcommand{\arraystretch}{1.5}
\begin{tabular*}{\linewidth}
{@{\extracolsep{\fill}}lcccccc@{}}
\hline\hline
 & \multicolumn{2}{c}{Discrimination} & \multicolumn{4}{c}{Signal efficiency at $\epsB$} \\
\cline{2-3}\cline{4-7}
Configuration & AUC & Max. SIC & $10\%$ & $5\%$ & $1\%$ & $0.5\%$ \\
\hline
\RIDDLE{} &
$0.898^{+0.02}_{-0.02}$ &
$14.90^{+0.13}_{-0.21}$ &
$0.779^{+0.03}_{-0.03}$ &
$0.729^{+0.03}_{-0.03}$ &
$0.606^{+0.03}_{-0.04}$ &
$0.532^{+0.03}_{-0.10}$ \\
IAD &
$0.890^{+0.01}_{-0.01}$ &
$14.46^{+0.76}_{-0.50}$ &
$0.780^{+0.01}_{-0.01}$ &
$0.726^{+0.01}_{-0.01}$ &
$0.606^{+0.01}_{-0.02}$ &
$0.547^{+0.01}_{-0.02}$ \\
Supervised &
$0.968^{+0.00}_{-0.00}$ &
$14.61^{+0.46}_{-0.30}$ &
$0.922^{+0.00}_{-0.00}$ &
$0.875^{+0.00}_{-0.00}$ &
$0.706^{+0.00}_{-0.00}$ &
$0.611^{+0.00}_{-0.00}$ \\
LaCATHODE &
$0.759^{+0.04}_{-0.06}$ &
$7.17^{+6.23}_{-2.41}$ &
$0.539^{+0.08}_{-0.07}$ &
$0.481^{+0.07}_{-0.05}$ &
$0.346^{+0.10}_{-0.05}$ &
$0.296^{+0.11}_{-0.05}$ \\[2pt]
\hline\hline
\end{tabular*}
\end{table*}

\begin{figure*}[!t]
\centering
\plotpanel{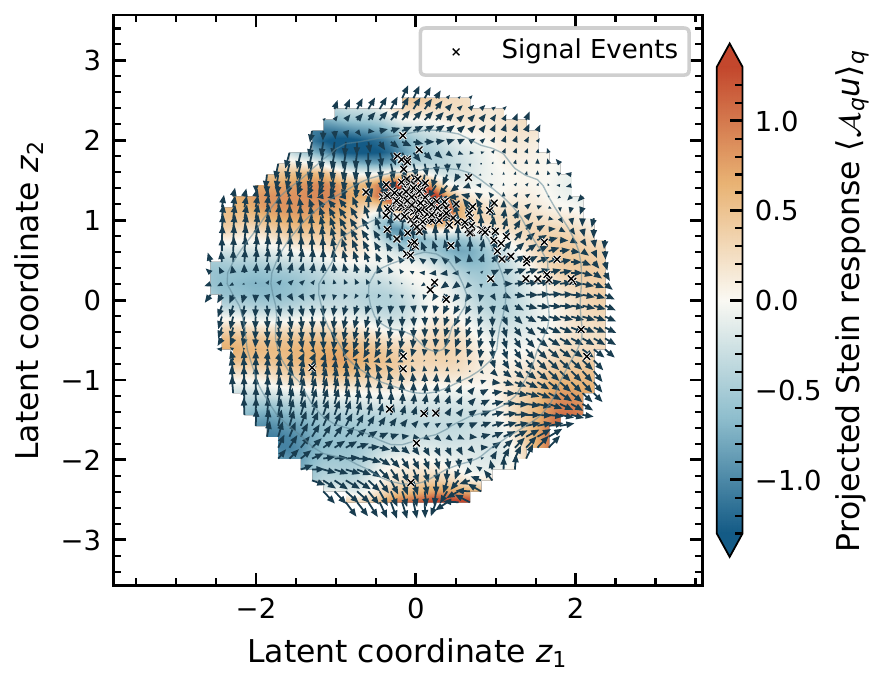}{(a) Signal-injected SR}\hfill%
\plotpanel{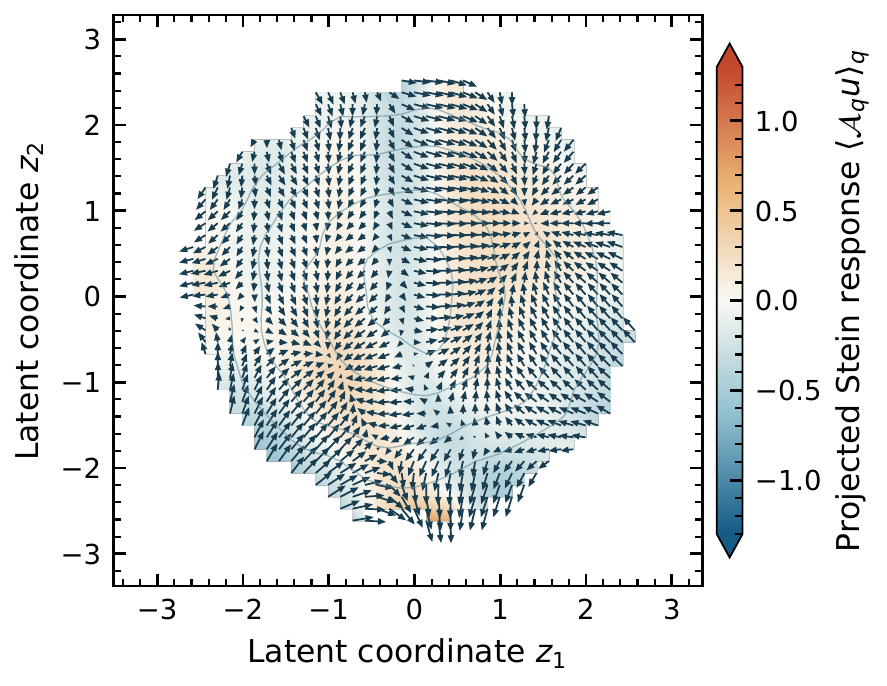}{(b) Background-only SR}
\caption{Projected \RIDDLE{} Stein responses for the mass-shifted observables, using witnesses trained separately on (a) signal-injected and (b) background-only SR samples. Colors show binned, smoothed averages of $\mathcal A_q u_\theta$ over generated-background reference events and the witness ensemble; arrows show corresponding averages of $-\nabla_{\zb}u_\theta$, indicating directions of decreasing witness values. Both are evaluated in four-dimensional latent space before projection onto $(z_1,z_2)$.}
\label{fig:shifted-witness}
\end{figure*}

Figure~\ref{fig:shifted-witness} compares two \RIDDLE{} witnesses trained separately on the signal-injected and background-only SR samples. As in Figures~\ref{fig:main-witness} and~\ref{fig:deltar-witness}, localized structure in the signal-injected response overlaps the signal population. The signal positions are added using simulation truth only after training and are not used to learn the witness. The shifted-feature projection also contains broader regions of positive response extending beyond the main signal concentration. In the latent representation learned from the shifted inputs, the Stein response combines the witness's local curvature with the alignment of its gradient and the background density score. Changes in their balance can produce positive responses beyond the main signal concentration, alongside localized structure overlapping the signal events. As in the original four-observable and $\Delta R$ analyses in Secs.~\ref{sec:mainresults} and~\ref{sec:deltar}, event selection uses the final calibrated anomaly score rather than the projected Stein response. Despite these differences in the projected response, the final signal retention and background rejection remain comparable to those obtained with the original four observables, as shown in Tables~\ref{tab:perf-main} and~\ref{tab:perf-shifted}.

\begin{figure*}[!t]
\centering
\plotpanel{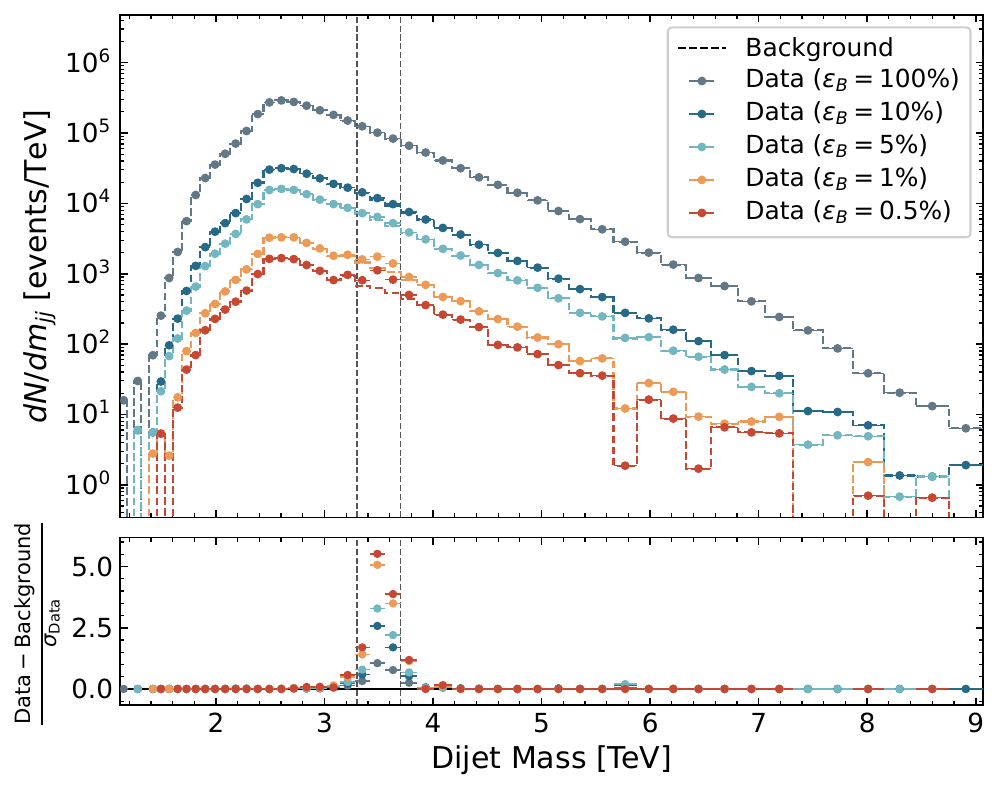}{}\hfill%
\plotpanel{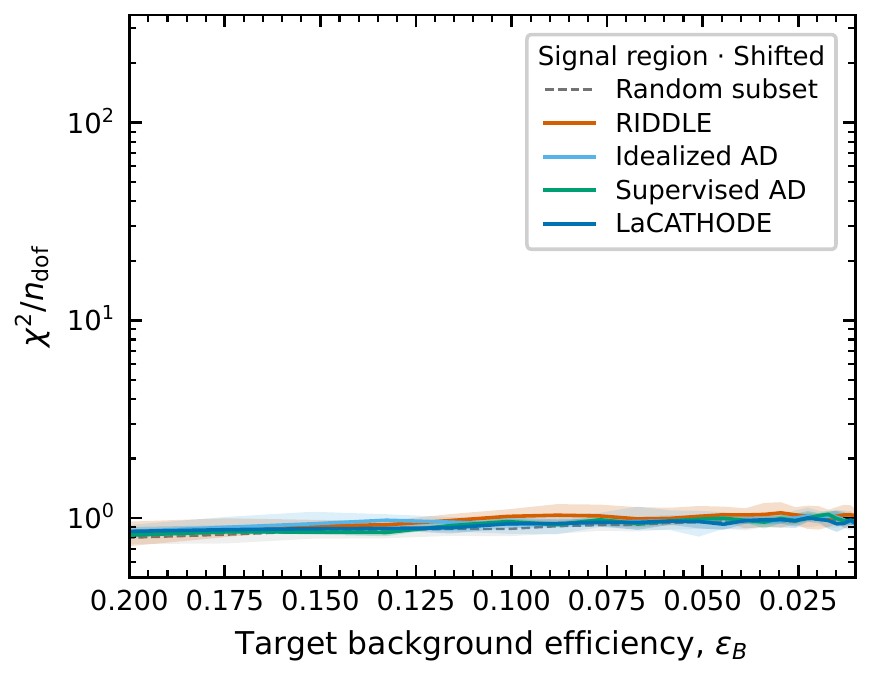}{}
\caption{Dijet-mass spectrum and background mass-sculpting test using the mass-shifted observables. Left: signal-injected \RIDDLE{} spectrum before and after selection at the indicated background acceptances, using the same score thresholds in the SR and sidebands. Right: background-only SR $\chi^2/n_{\rm dof}$ for \RIDDLE{}, IAD, Supervised, and LaCATHODE. Colored curves show means over five runs, with shaded bands spanning the run minima and maxima. The gray dashed line shows the median of 100 random QCD selections at matching background efficiencies.}
\label{fig:mass-shifted}
\end{figure*}

Figure~\ref{fig:mass-shifted} shows the dijet-mass spectrum and background mass-sculpting test using the same evaluation procedure as in the original and $\Delta R$ analyses. The same scoring procedure and thresholds are applied in the SR and sidebands. The spectrum again shows that tighter score requirements suppress the QCD background relative to the signal peak in the SR. In the mass-sculpting test, all four methods yield background-only $\chi^2/n_{\rm dof}$ values close to the random-selection reference across the tested background efficiencies, indicating that the SR QCD mass shape remains largely stable despite the imposed mass dependence. Together with the discrimination results, this shows that \RIDDLE{} maintains effective background rejection and signal retention with shifted observables.

\resultsectionbreak
\section{Sensitivity at lower signal injection}\label{sec:injection}

The sensitivity of \RIDDLE{} to lower signal fractions is examined by reducing the number of injected signal events while retaining the original four observables. Following Ref.~\cite{hallin2022cathode}, the study compares the maximum SIC and the approximate counting sensitivity defined in Sec.~\ref{subsec:metrics}. The scan covers $N_{\mathrm{inj}}=1000,667,500,333,250,167,83$, where $N_{\mathrm{inj}}$ denotes the number of signal events drawn from the prepared reservoir before applying the SR mass requirement. The background-sample construction and held-out evaluation population remain unchanged. The injected signal fraction is varied in the SR mixture used to train \RIDDLE{}, IAD, and LaCATHODE, while Supervised retains the independent known-signal training sample described in Sec.~\ref{subsec:benchmarks}.

The study also examines the dependence of discrimination on witness-training duration and checkpoint selection. The Stein-witness results in Secs.~\ref{sec:mainresults}--\ref{sec:shifted} use the ten checkpoints with the lowest validation Stein loss in each fit (Appendix~\ref{app:numerical}). For the injection scan, an alternative strategy uses the checkpoint at epoch 200 from each fit and averages the event scores across all twenty fits. Additional training up to 300 epochs showed no appreciable improvement beyond 200 epochs. The score construction and calibration follow the same procedure for both checkpoint-selection strategies.

\begin{figure*}[!t]
\centering
\plotpanel{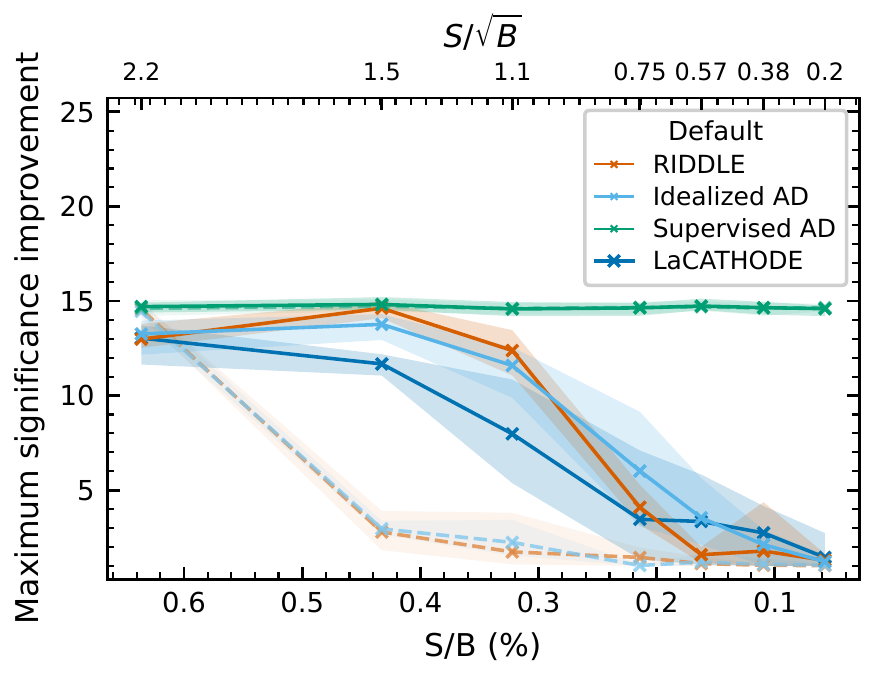}{}\hfill%
\plotpanel{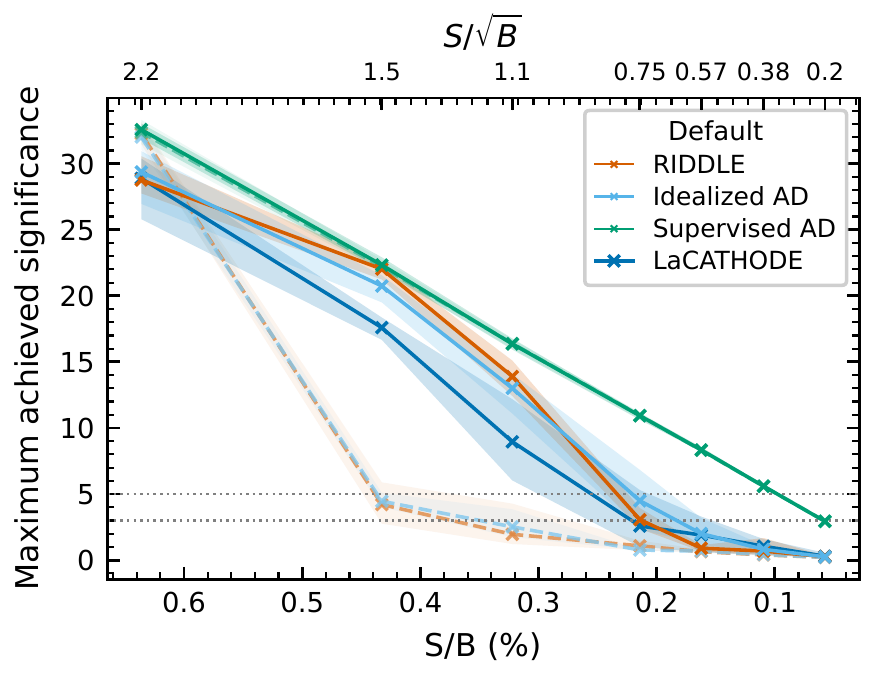}{}
\caption{Signal-injection dependence using the original four observables. Left: maximum SIC. Right: maximum approximate counting sensitivity $Z_{\mathrm{count}}$. For \RIDDLE{}, IAD, and Supervised, solid lines use the checkpoint at epoch 200 from each fit, while dashed lines use the ten checkpoints with the lowest validation Stein loss. LaCATHODE follows its original checkpoint-selection procedure. The lower horizontal axis shows the signal-to-background ratio in the SR, and the upper axis gives the corresponding $S/\sqrt B$ before anomaly-score selection. Points show means over five runs with different random seeds, with bands spanning the run minima and maxima.}
\label{fig:injection}
\end{figure*}

The left panel of Figure~\ref{fig:injection} shows how the maximum SIC changes as the injected signal fraction decreases. At the reference injection, \RIDDLE{} achieves a maximum SIC comparable to IAD and Supervised without using signal labels during training. Supervised maintains a nearly constant maximum SIC because its known-signal training population remains unchanged throughout the scan. In contrast, the discrimination achieved by \RIDDLE{}, IAD, and LaCATHODE decreases as the signal fraction in the unlabeled SR mixture becomes smaller.

For \RIDDLE{} and IAD, selecting the checkpoint at epoch 200 yields substantially higher maximum SIC than selecting checkpoints by the lowest validation Stein loss at several reduced injection strengths. This indicates that a lower validation Stein loss does not necessarily lead to better discrimination of rare signal events. With the fixed-epoch strategy, \RIDDLE{} achieves higher mean maximum SIC than LaCATHODE at intermediate injection strengths and comparable performance at the lowest signal fraction.

The right panel of Figure~\ref{fig:injection} shows the maximum approximate counting sensitivity $Z_{\mathrm{count}}$, calculated from the maximum SIC and the uncut SR $S/\sqrt B$, as defined in Sec.~\ref{subsec:metrics}. Unlike SIC alone, this measure also accounts for the signal yield in the SR. As fewer signal events are injected, the counting sensitivity decreases even for Supervised, whose maximum SIC remains nearly constant. Nevertheless, the improvement obtained with the fixed-epoch strategy extends the sensitivity of \RIDDLE{} to lower signal fractions.

These results motivate considering both checkpoint-selection strategies. While validation-based selection remains effective at the reference injection, extended training with fixed-epoch selection can improve discrimination at lower signal fractions without requiring signal labels.

\Needspace*{20\baselineskip}
\section{Conclusions}\label{sec:conclusions}

This work introduced \RIDDLE{}, a method for resonant anomaly detection that learns a scalar Stein-witness function from unlabeled signal-region events. The witness identifies discrepancies in a latent representation of auxiliary jet observables relative to a background reference inferred from dijet-mass sidebands. This allows events to be ranked without requiring an explicit signal model or truth labels during training.

In the LHC Olympics study, \RIDDLE{} achieves strong signal--background discrimination and performs close to IAD at the reference signal injection, despite estimating the background reference from sidebands rather than known background events in the signal region. Compared with LaCATHODE, \RIDDLE{} achieves higher mean AUC, maximum SIC, and signal efficiencies across the three studied feature configurations, with comparable background mass-shape stability.

\RIDDLE{} also retains effective discrimination when interjet angular separation is included and when an explicit mass dependence is introduced into the jet-mass observables. The improvement relative to LaCATHODE is particularly pronounced with the mass-shifted observables, for which \RIDDLE{} achieves more than twice the mean maximum SIC. Across the studied configurations, anomaly-score selection makes the injected signal peak more pronounced relative to the QCD background without appreciable background mass sculpting.

The signal-injection study further shows that extended witness training and fixed-epoch checkpoint selection can improve discrimination at reduced signal fractions. These results motivate examining both validation-based and fixed-epoch checkpoint-selection strategies for rare-signal searches. Together, the results support the robustness of \RIDDLE{} under the studied conditions and its potential for signal enrichment in resonant searches.
\begin{acknowledgments}
I thank the developers and contributors of the LHC Olympics samples for making the simulated events publicly available~\cite{kasieczka2021lhco}. I am also grateful to the members of the High Energy Physics group at The Catholic University of America for fostering a welcoming and supportive academic environment. This work used resources available through the National Research Platform (NRP)~\cite{weitzel2025nrp} at the University of California, San Diego. NRP has been developed, and is supported in part, by funding from the National Science Foundation, from awards 1730158, 1540112, 1541349, 1826967, 2112167, 2100237, and 2120019, as well as additional funding from community partners. 

\end{acknowledgments}

\section*{Data and RIDDLE Framework Availability}
The \RIDDLE{} framework, including the analysis and plotting code, is publicly available at \url{https://github.com/asimsek/RIDDLE}. The simulated dijet events used in this study are available from the LHC Olympics R\&D dataset~\cite{kasieczka2019lhcord}. The additional QCD background sample is available separately~\cite{shih2021additionalqcd}.

\appendix

\section{The conditional Stein identity}\label{app:stein}

At fixed dijet mass, let $q_m(\zb)=q(\zb\mid m)$ denote the background reference and let $\rho_m(\zb)$ denote another density on the latent domain $\Omega$. The densities are taken to be positive and differentiable within $\Omega$, and the witness $u$ has first and second derivatives. The averages below are assumed to be finite. The vector field $\bm f(\zb)=\nabla_{\zb}u(\zb)$ is the gradient of the scalar witness.

The mean response under the background reference is obtained by weighting the response at each point in latent space by the background density:
\begin{equation}
\mathbb E_{q_m}[\mathcal A_q u]
=
\int_{\Omega}q_m(\zb)\,
\mathcal A_q u(\zb,m)\,\dd^d\zb.
\label{eq:backgroundaverage}
\end{equation}
For an average under $\rho_m$, integration by parts on the Laplacian term, $\Delta_{\zb}u=\nabla_{\zb}\!\cdot\bm f$, gives
\begin{align}
\mathbb E_{\rho_m}[\mathcal A_q u]
&=\int_{\partial\Omega}\rho_m\bm f\cdot\bm n\,\dd A
\nonumber\\
&\quad+\mathbb E_{\rho_m}\!\left[
\bm f\cdot\nabla_{\zb}\log\frac{q_m}{\rho_m}
\right].
\label{eq:integration}
\end{align}
Here, $\partial\Omega$ denotes the boundary of the latent domain, $\bm n$ its outward unit normal, and $\dd A$ the surface element. For the zero-mean identity, this boundary integral must vanish. For an unbounded domain, this means that the surface integral must approach zero as the boundary moves to infinity.

When $\rho_m=q_m$, the ratio inside the logarithm is one, so the second term in Eq.~\eqref{eq:integration} is zero. If the boundary integral also vanishes, the mean Stein response is therefore zero, as given in Eq.~\eqref{eq:steinidentity}. Equivalently, the two terms in the Stein response make equal and opposite contributions to the background average.

For the observed density $\rho_m=p(\zb\mid m)$, the logarithmic gradient instead becomes
$\nabla_{\zb}\log q_m-\nabla_{\zb}\log p(\zb\mid m)$.
With the same boundary condition, the expected response is determined by this difference between the background and data density scores, together with the witness gradient. A suitable witness can therefore reveal a mismatch between the two distributions~\cite{grathwohl2020stein,repasky2023critic}.

The identity holds separately at each fixed mass. During training, a single witness is used throughout the SR, and the response is averaged over the joint distribution $p(\zb,m)$:
\begin{equation}
\mathbb E_p[\mathcal A_q u]
=
\int_{\mathrm{SR}} p(m)\,
\mathbb E_{p(\zb\mid m)}[\mathcal A_q u(\zb,m)]\,\dd m.
\label{eq:massaverage}
\end{equation}
Here, $p(m)$ is the mass distribution of the SR training population. The conditional mean responses are therefore weighted according to the occurrence of each mass in that population.

The witness is optimized to reveal differences between the signal-region distribution and the background reference rather than to directly estimate the log-density ratio. In \RIDDLE{}, the same neural-network witness is used across the SR mass range and is trained with the regularized Stein loss and the additional tail-ranking term. Its output and derivatives are then used to construct the calibrated anomaly score. This score ranks events relative to the background reference and should not be interpreted as an exact likelihood ratio.


\section{Numerical estimator and calibration}\label{app:numerical}

The witness is a neural network with three hidden layers of 128 SiLU~\cite{elfwing2018silu} units each and a single scalar output. It receives only latent coordinates, while the background reference depends on dijet mass. Training uses AdamW~\cite{loshchilov2019adamw} with 256 events per minibatch, a learning rate of $3\times10^{-4}$, and weight decay $10^{-2}$. The gradient-penalty strength in the principal Stein loss is fixed at $\lambda=1$. Twenty fits are trained on the same prepared event population. For the Stein-witness results in Secs.~\ref{sec:mainresults}--\ref{sec:shifted}, event scores from the ten checkpoints with the lowest validation Stein loss are averaged within each fit. Section~\ref{sec:injection} also examines an alternative checkpoint-selection strategy that uses only the checkpoint at epoch 200 from each fit.

The auxiliary tail-ranking loss focuses on separating high-scoring training events from background events that also receive high scores. For this comparison, events are ranked using the negative of the learned scalar witness output, $P=-u_\theta$. This sign convention is motivated by the optimal regularized vector critic~\cite{hu2018stein,grathwohl2020stein}: at fixed mass, the idealized scalar witness has a gradient proportional to $\nabla_{\zb}\log[q(\zb\mid m)/p(\zb\mid m)]$. Under this relation, regions with a larger data density relative to the background have lower witness values. Taking the negative, $P=-u_\theta$, therefore gives higher scores to regions where the data density is enhanced relative to the expected background density. The idealized density-ratio relation motivates this sign convention, but the trained witness need not satisfy it exactly. The sign of the witness alone does not identify signal or background. Only the ordering matters: an event with a lower witness value receives a higher ranking score. To compare events at similar dijet masses, each minibatch is ordered by mass and divided into eight groups of approximately equal size. The highest-scoring 10\% of training events are selected within each group, allowing the auxiliary loss to focus on potentially anomalous events rather than the background-dominated bulk of the training sample.

The background comparison is drawn from a larger generated sample. For every training event, eight background events are generated at the same dijet mass, giving a candidate pool eight times the minibatch size. Within each mass group, the highest-scoring 1\% of these background candidates are selected. This focuses the comparison on background events that are difficult to reject using the current witness ranking, rather than on typical low-scoring background events. Any fractional selection count is rounded up to the next whole event. The selected training and background samples do not need to contain equal numbers of events.

Each selected training event, with score $P_i=-u_\theta(\zb_i)$, is compared with every background event retained in the highest-scoring 1\% of the same mass group. The background score $P_j^{\mathrm{bg}}=-u_\theta(\zb_j^{\mathrm{bg}})$ is calculated using the same witness network. For each such pair, the pairwise logistic penalty~\cite{burges2005learning} is
\begin{equation}
\ell_{ij}
=
\log\!\left[1+\exp\!\left(P_j^{\mathrm{bg}}-P_i\right)\right].
\label{eq:tailpair}
\end{equation}
The penalty depends directly on the difference between the two scores. It is larger when the background event receives the higher score and decreases as the training-event score increases relative to the background-event score.

For each selected training event, the pairwise penalties are averaged over the selected background events in the same mass group. Averaging these values over all selected training events gives $\mathcal L_{\mathrm{tail}}$. Let $\mathcal I_g$ and $\mathcal B_g$ denote the sets of selected training and generated-background events in mass group $g$, respectively. The auxiliary loss is
\begin{equation}
\mathcal L_{\mathrm{tail}}
=
\frac{1}{N_{\mathrm{sel}}}
\sum_{g=1}^{8}
\sum_{i\in\mathcal I_g}
\left[
\frac{1}{|\mathcal B_g|}
\sum_{j\in\mathcal B_g}\ell_{ij}
\right],
\label{eq:tailloss}
\end{equation}
where $|\mathcal B_g|$ is the number of selected background events in group $g$, and $N_{\mathrm{sel}}=\sum_{g=1}^{8}|\mathcal I_g|$ is the total number of selected training events. Each selected training event therefore contributes equally to the loss, even when the groups contain different numbers of selected background events.

In \RIDDLE{}, the SR events entering this loss are selected by their witness scores, without using their true signal or background labels. The selected sample can therefore contain both signal and background. Training uses only the principal Stein loss for the first five epochs. From the sixth epoch onward, the training loss is $\mathcal L_{\mathrm{Stein}}+0.005\,\mathcal L_{\mathrm{tail}}$. The validation loss remains $\mathcal L_{\mathrm{Stein}}$ alone. Under validation-based checkpoint selection, the ten checkpoints with the lowest validation loss are selected in each fit.

For final score construction, a separate sample of 524,288 generated-background events is randomly divided into two equally sized reference subsets, A and B, each containing 262,144 events. The score is built from three quantities derived from the witness: the reversed witness output, $P=-u_\theta$; the sensitivity of the witness to small changes in the latent coordinates, measured by $E=\|\nabla_{\zb}u_\theta\|^2$; and the Stein response, $W=\mathcal A_q u_\theta$. Subset A provides the background reference values used to normalize these quantities, allowing them to be combined on a common scale. Subset B is used to determine a support correction, $\Pi(\zb,m)$, which reduces scores for events with unusually large latent distances or large gradients of the background log density. The corrected scores from subset B are also used for the final calibration.

For normalization, subset A is ordered by dijet mass and divided into eight groups containing approximately equal numbers of events. Within each group, the three quantities $P$, $E$, and $W$ are compared separately with their corresponding background values to determine their percentiles. These percentiles are then interpolated between the mass-group centers according to the event's dijet mass. Each percentile is converted to the value at the same percentile of a standard normal distribution: the median maps to zero, values above the median become positive, and values below it become negative. The normalized quantities, $\widetilde P$, $\widetilde E$, and $\widetilde W$, therefore describe how each component compares with its own background distribution on a common numerical scale.

For each retained witness network, the normalized quantities are combined into a score:
\begin{equation}
R_\theta=\widetilde P+1.5\widetilde E
+0.6\log\!\left[
1+\exp\!\left(\frac{\widetilde W-1.96}{0.25}\right)
\right].
\label{eq:rawscore}
\end{equation}
Here, the coefficients $1.5$ and $0.6$ are fixed scoring hyperparameters that determine the contributions of the gradient-energy and Stein-response terms, respectively. The $1.96$ corresponds to the 97.5th percentile of a standard normal distribution, while the fixed parameter $0.25$ controls how rapidly the Stein-response term increases around this threshold. This term contributes little when $\widetilde W$ is well below $1.96$ and increases smoothly for larger values, giving greater weight to Stein responses that are unusually high relative to the background. This helps prioritize events that may deviate from the expected background distribution. Under validation-based checkpoint selection, the scores from the ten retained networks are averaged with equal weight within each fit. Under fixed-epoch selection, each fit contributes the score from a single checkpoint. The fit-level scores are then averaged equally across the fits to obtain $R_{\rm ensemble}$.

A support correction is then applied to reduce scores for events with properties that are unusually extreme relative to the generated background. This correction uses two measures: the distance from the origin in latent space, $\|\zb\|$, and the magnitude of the background log-density gradient, $\|\bm s_q\|$, which measures how rapidly the background log density changes near the event. Each measure is compared separately with its corresponding distribution in reference subset B, divided into eight dijet-mass groups. The resulting percentiles are converted to standard-normal values using the same procedure described above. Large normalized values therefore indicate an unusually large latent distance or a rapidly changing background density relative to background events at similar dijet masses.

For each of these two measures, a smooth penalty is calculated from its normalized value. Let $\xi_r$ and $\xi_s$ denote the normalized latent-distance and background-gradient measures, respectively. The penalty is calculated using
\begin{equation}
h(\xi)=2\log\!\left[
1+\exp\!\left(
\frac{\xi-\Phi^{-1}(0.999)}{0.25}
\right)\right].
\label{eq:penaltyfunction}
\end{equation}
Here, $\Phi^{-1}$ is the inverse cumulative distribution function of a standard normal distribution, and $\Phi^{-1}(0.999)\simeq 3.09$ corresponds to its 99.9th percentile. The factor $2$ sets the overall strength of the penalty, while the fixed parameter $0.25$ controls how rapidly it increases near this threshold. The support correction is then defined as the larger of the two penalties:
\begin{equation}
\Pi(\zb,m)=\max\{h(\xi_r),h(\xi_s)\}.
\label{eq:supportpenalty}
\end{equation}
The support correction reduces $R_{\rm ensemble}$, particularly for events that lie far from the origin in latent space or where the background log density changes rapidly. Such events may be more sensitive to errors in the background model, so their scores are reduced rather than the events being rejected outright. 

The final calibration compares each event's corrected score with the scores of reference background events at similar dijet masses. To make this comparison, the reference events in subset B are also scored and corrected in the same way as the events being evaluated. Let $r=R_{\rm ensemble}(\zb,m)-\Pi(\zb,m)$ denote an event's corrected score. The function $F_q(r\mid m)$ gives the fraction of reference background events at similar dijet masses with corrected scores at or below $r$. For this calibration, subset B is ordered by dijet mass and divided into two equally sized groups. The background percentiles are calculated separately within each group and linearly interpolated between their mean dijet masses. The final event score is
\begin{equation}
S(\zb,m)=
\bigl[
F_q(R_{\rm ensemble}(\zb,m)-\Pi(\zb,m)\mid m)
\bigr]^{10}.
\label{eq:fullscore}
\end{equation}

An event receives a high final score when its corrected score is higher than those of most reference background events at similar dijet masses. This indicates how unusual the event is relative to the background, rather than the probability of a signal. Raising the background percentile to the power ten gives greater emphasis to high percentile values without changing their ordering. However, the support correction and mass-dependent calibration can change the original ranking of events based on $R_{\rm ensemble}$. The background reference distributions and calibration rules are fixed before the held-out events are evaluated.

For the comparison with LaCATHODE~\cite{hallin2023lacathode}, the background flow and classifier are trained for 100 epochs each, following the configuration of the public implementation. Classifier predictions are averaged over the ten checkpoints with the lowest validation loss. The comparison workflow, including the implementation version and evaluation procedure, is provided in the \RIDDLE{} repository.

\Needspace{6\baselineskip}
\section{Background-only Stein-witness populations}\label{app:bgpopulations}

The background-only study uses separate event samples for background modeling, witness training, and evaluation. Unlike the signal-injected analyses summarized in Table~\ref{tab:populations}, this study contains no injected signal events in the ordinary SR or sideband samples, so some training and validation counts are slightly smaller. Nevertheless, both studies use the same underlying background population, with the differences in event counts arising from the injected signal events. The sideband mapping uses 307,311 training and 73,419 validation events. The \RIDDLE{} background correction uses separate samples containing 131,705 training and 36,710 validation events. The \RIDDLE{} and IAD witnesses are trained on 48,787 QCD-only SR events, with 12,197 events reserved for validation. Supervised instead retains its independent sample of 29,810 known-signal training events and 7,453 validation events. The background-only designation therefore refers to the ordinary SR population and the evaluation sample, not to the independent known-signal sample used for Supervised training. The held-out background-only SR sample contains 386,472 QCD events. The independent pure-background reference used by IAD and Supervised retains the population sizes and roles summarized in Table~\ref{tab:populations}.

\FloatBarrier
\bibliography{references}

@article{kasieczka2021lhco,
  author = {Kasieczka, Gregor and Nachman, Benjamin and Shih, David and others},
  title = {The {LHC} Olympics 2020: A Community Challenge for Anomaly Detection in High Energy Physics},
  journal = {Reports on Progress in Physics},
  volume = {84},
  pages = {124201},
  year = {2021},
  doi = {10.1088/1361-6633/ac36b9},
  eprint = {2101.08320},
  archivePrefix = {arXiv}
}

@misc{kasieczka2019lhcord,
  author = {Kasieczka, Gregor and Nachman, Ben and Shih, David},
  title = {{R\&D Dataset for LHC Olympics 2020 Anomaly Detection Challenge}},
  howpublished = {Zenodo, version 5},
  year = {2019},
  doi = {10.5281/zenodo.6466204}
}

@misc{shih2021additionalqcd,
  author = {Shih, David},
  title = {{Additional QCD Background Events for LHCO2020 R\&D (signal region only)}},
  howpublished = {Zenodo, version 1},
  year = {2021},
  doi = {10.5281/zenodo.5759087}
}

@article{nachman2020anode,
  author = {Nachman, Benjamin and Shih, David},
  title = {Anomaly Detection with Density Estimation},
  journal = {Phys. Rev. D},
  volume = {101},
  pages = {075042},
  year = {2020},
  doi = {10.1103/PhysRevD.101.075042},
  eprint = {2001.04990},
  archivePrefix = {arXiv}
}

@article{hallin2022cathode,
  author = {Hallin, Anna and Isaacson, Joshua and Kasieczka, Gregor and Krause, Claudius and Nachman, Benjamin and Quadfasel, Tobias and Schlaffer, Matthias and Shih, David and Sommerhalder, Manuel},
  title = {Classifying Anomalies Through Outer Density Estimation},
  journal = {Phys. Rev. D},
  volume = {106},
  pages = {055006},
  year = {2022},
  doi = {10.1103/PhysRevD.106.055006},
  eprint = {2109.00546},
  archivePrefix = {arXiv}
}

@article{hallin2023lacathode,
  author = {Hallin, Anna and Kasieczka, Gregor and Quadfasel, Tobias and Shih, David and Sommerhalder, Manuel},
  title = {Resonant Anomaly Detection without Background Sculpting},
  journal = {Phys. Rev. D},
  volume = {107},
  pages = {114012},
  year = {2023},
  doi = {10.1103/PhysRevD.107.114012},
  eprint = {2210.14924},
  archivePrefix = {arXiv}
}

@inproceedings{grathwohl2020stein,
  author = {Grathwohl, Will and Wang, Kuan-Chieh and Jacobsen, J{\o}rn-Henrik and Duvenaud, David and Zemel, Richard},
  title = {Learning the {Stein} Discrepancy for Training and Evaluating Energy-Based Models without Sampling},
  booktitle = {Proceedings of the 37th International Conference on Machine Learning},
  volume = {119},
  pages = {3732--3747},
  publisher = {PMLR},
  year = {2020},
  eprint = {2002.05616},
  archivePrefix = {arXiv}
}

@article{repasky2023critic,
  author = {Repasky, Matthew and Cheng, Xiuyuan and Xie, Yao},
  title = {Neural {Stein} Critics with Staged {$L^2$}-Regularization},
  journal = {IEEE Trans. Inf. Theory},
  volume = {69},
  pages = {7246--7275},
  year = {2023},
  doi = {10.1109/TIT.2023.3299258},
  eprint = {2207.03406},
  archivePrefix = {arXiv}
}

@article{hyvarinen2005score,
  author = {Hyv{\"a}rinen, Aapo},
  title = {Estimation of Non-Normalized Statistical Models by Score Matching},
  journal = {J. Mach. Learn. Res.},
  volume = {6},
  pages = {695--709},
  year = {2005},
  url = {https://jmlr.org/papers/v6/hyvarinen05a.html}
}

@article{cowan2011asymptotic,
  author = {Cowan, Glen and Cranmer, Kyle and Gross, Eilam and Vitells, Ofer},
  title = {Asymptotic Formulae for Likelihood-Based Tests of New Physics},
  journal = {Eur. Phys. J. C},
  volume = {71},
  pages = {1554},
  year = {2011},
  doi = {10.1140/epjc/s10052-011-1554-0},
  eprint = {1007.1727},
  archivePrefix = {arXiv}
}

@inproceedings{weitzel2025nrp,
  author = {Weitzel, Derek and Graves, Ashton and Albin, Sam and Zhu, Huijun and Wuerthwein, Frank and Tatineni, Mahidhar and Mishin, Dmitry and Khoda, Elham and Sada, Mohammad and Smarr, Larry and DeFanti, Thomas and Graham, John},
  title = {The National Research Platform: Stretched, Multi-Tenant, Scientific Kubernetes Cluster},
  booktitle = {Practice and Experience in Advanced Research Computing 2025},
  publisher = {ACM},
  year = {2025},
  doi = {10.1145/3708035.3736060},
  eprint = {2505.22864},
  archivePrefix = {arXiv}
}

@inproceedings{stein1972bound,
  title = {A Bound for the Error in the Normal Approximation to the Distribution of a Sum of Dependent Random Variables},
  author = {Stein, Charles},
  booktitle = {Proceedings of the Sixth Berkeley Symposium on Mathematical Statistics and Probability, Volume II: Probability Theory},
  pages = {583--602},
  publisher = {University of California Press},
  year = {1972},
  url = {https://digicoll.lib.berkeley.edu/record/113245}
}

@article{thaler2011nsubjettiness,
  author = {Thaler, Jesse and Van Tilburg, Ken},
  title = {Identifying Boosted Objects with {$N$}-subjettiness},
  journal = {Journal of High Energy Physics},
  volume = {03},
  pages = {015},
  year = {2011},
  doi = {10.1007/JHEP03(2011)015},
  eprint = {1011.2268},
  archivePrefix = {arXiv}
}

@article{black2011sic,
  author = {Gallicchio, Jason and Huth, John and Kagan, Michael and Schwartz, Matthew D. and Black, Kevin and Tweedie, Brock},
  title = {Multivariate Discrimination and the {Higgs} + {W}/{Z} Search},
  journal = {Journal of High Energy Physics},
  volume = {04},
  pages = {069},
  year = {2011},
  doi = {10.1007/JHEP04(2011)069},
  eprint = {1010.3698},
  archivePrefix = {arXiv}
}

@article{cacciari2008antikt,
  author = {Cacciari, Matteo and Salam, Gavin P. and Soyez, Gregory},
  title = {The Anti-{$k_t$} Jet Clustering Algorithm},
  journal = {Journal of High Energy Physics},
  volume = {04},
  pages = {063},
  year = {2008},
  doi = {10.1088/1126-6708/2008/04/063},
  eprint = {0802.1189},
  archivePrefix = {arXiv}
}

@inproceedings{loshchilov2019adamw,
  author = {Loshchilov, Ilya and Hutter, Frank},
  title = {Decoupled Weight Decay Regularization},
  booktitle = {International Conference on Learning Representations},
  year = {2019},
  eprint = {1711.05101},
  archivePrefix = {arXiv}
}

@article{elfwing2018silu,
  author = {Elfwing, Stefan and Uchibe, Eiji and Doya, Kenji},
  title = {Sigmoid-Weighted Linear Units for Neural Network Function Approximation in Reinforcement Learning},
  journal = {Neural Netw.},
  volume = {107},
  pages = {3--11},
  year = {2018},
  doi = {10.1016/j.neunet.2017.12.012},
  eprint = {1702.03118},
  archivePrefix = {arXiv}
}

@misc{hu2018stein,
  author = {Hu, Tianyang and Chen, Zixiang and Sun, Hanxi and Bai, Jincheng and Ye, Mao and Cheng, Guang},
  title = {Stein Neural Sampler},
  year = {2018},
  eprint = {1810.03545},
  archivePrefix = {arXiv}
}

@inproceedings{gorham2015stein,
  author = {Gorham, Jackson and Mackey, Lester},
  title = {Measuring Sample Quality with {Stein's} Method},
  booktitle = {Advances in Neural Information Processing Systems},
  volume = {28},
  pages = {226--234},
  year = {2015},
  eprint = {1506.03039},
  archivePrefix = {arXiv}
}

@article{collins2018cwola,
  author = {Collins, Jack H. and Howe, Kiel and Nachman, Benjamin},
  title = {Anomaly Detection for Resonant New Physics with Machine Learning},
  journal = {Phys. Rev. Lett.},
  volume = {121},
  pages = {241803},
  year = {2018},
  doi = {10.1103/PhysRevLett.121.241803},
  eprint = {1805.02664},
  archivePrefix = {arXiv}
}

@article{raine2023curtains,
  author = {Raine, John Andrew and Klein, Samuel and Sengupta, Debajyoti and Golling, Tobias},
  title = {{CURTAINs} for Your Sliding Window: Constructing Unobserved Regions by Transforming Adjacent Intervals},
  journal = {Front. Big Data},
  volume = {6},
  pages = {899345},
  year = {2023},
  doi = {10.3389/fdata.2023.899345},
  eprint = {2203.09470},
  archivePrefix = {arXiv}
}

@inproceedings{burges2005learning,
  author = {Burges, Christopher J. C. and Shaked, Tal and Renshaw, Erin and Lazier, Ari and Deeds, Matt and Hamilton, Nicole and Hullender, Gregory N.},
  title = {Learning to Rank Using Gradient Descent},
  booktitle = {Proceedings of the 22nd International Conference on Machine Learning},
  pages = {89--96},
  publisher = {ACM},
  year = {2005},
  doi = {10.1145/1102351.1102363}
}
\end{document}